\documentclass[aps,prd,10pt,onecolumn,preprintnumbers,superscriptaddress,showpacs,floatfix,nofootinbib]{revtex4-2}

\usepackage[utf8]{inputenc}
\usepackage{textcomp,gensymb}
\usepackage{hyperref}
\usepackage{graphicx}
\usepackage{amsfonts,amsmath,amssymb,bm,bbm}
\usepackage{color,soul}
\usepackage[dvipsnames]{xcolor}
\usepackage{slashed}
\usepackage{flushend}
\usepackage{physics}
\usepackage{dcolumn}
\usepackage{comment}
\usepackage{tikz}
\usepackage[normalem]{ulem}
\usepackage[compat=1.1.0]{tikz-feynman}
\usepackage[caption=false]{subfig}

\hypersetup{ pdfnewwindow=true, colorlinks=true, linkcolor=blue, citecolor=blue, filecolor=blue, urlcolor=blue } 

\begin{document}
\begin{flushright}
\preprint{MI-HET-880}
\end{flushright}

\title{UV-Complete Models for a Light Axial Gauge Boson} 

\author{Bhaskar Dutta}
\email{dutta@tamu.edu}
\affiliation{
Mitchell Institute for Fundamental Physics and Astronomy,
Department of Physics and Astronomy, Texas A\&M University, College Station, TX 77843, USA
}
\author{Aparajitha Karthikeyan}
\email{aparajitha\_96@tamu.edu}
\affiliation{
Mitchell Institute for Fundamental Physics and Astronomy,
Department of Physics and Astronomy, Texas A\&M University, College Station, TX 77843, USA
}
\author{Rabindra N. Mohapatra}
\email{rmohapat@umd.edu}
\affiliation{
Maryland Center for Fundamental Physics \& Department of Physics, University of Maryland, College Park, Maryland 20742, USA
}

\date{\today}

\date{March 2026}

\begin{abstract}
    We present new anomaly free gauge models where the gauge field only has axial vector couplings to both quarks and leptons. We use the left-right symmetric universal seesaw models as the basis for this construction with an extra $U(1)_a$ as the axial gauge group. We present three main versions of the model, denoted as models A, B and C (and their variations), with different properties depending on the way the gauge anomaly is canceled. We show how the models can accommodate small neutrino masses. The models allow for a new Dirac fermion coupled via the $U(1)_a$ gauge portal to the SM fields which can be the dark matter. The models A and its variation have the novel property that there is an upper limit on the $U(1)_a$ gauge coupling $g_a$, due to the fact that the standard model Higgs doublet shares the  $U(1)_a$ quantum number. For models B and C, we discuss the phenomenological constraints on the gauge coupling $g_a$ and gauge boson mass $m_{\cal A}$ from current low energy observations where, unlike in models A, $g_a$ depends on $m_{\cal A}$  through a single vacuum expectation value. 
\end{abstract}

\maketitle

\section{Introduction} With LHC finding no new physics beyond the standard model (SM) in the TeV energy range and as we wait for higher energy and higher luminosity colliders to come online, attention has shifted to theoretical studies of possibilities of sub-GeV particles allowed by current data, which could have testable low energy manifestations. One such possibility is a low mass axial vector gauge boson which could play a role in the so-called ATOMKI observation~\cite{Feng:2016jff, Kozaczuk:2016nma, Barducci:2022lqd} of excited Beryllium decay~\cite{Krasznahorkay:2015iga}, $(g-2)$ of muon (for a review, see Ref.~\cite{Aliberti:2025beg}) as well as explanation of MiniBooNE low energy excess~\cite{MicroBooNE:2025ntu, MiniBooNE:2018esg} as discussed in, for example, Refs.~\cite{Bertuzzo:2018itn,Ballett:2018ynz, Fischer:2019fbw, Dentler:2019dhz, Dutta:2025fgz, Dutta:2021cip,Sorel:2003hf,Karagiorgi:2009nb,Collin:2016aqd,Giunti:2011gz,Giunti:2011cp,Gariazzo:2017fdh,Boser:2019rta,Kopp:2011qd,Kopp:2013vaa,Dentler:2018sju,Abazajian:2012ys,Conrad:2012qt,Diaz:2019fwt,Asaadi:2017bhx,Karagiorgi:2012kw,Pas:2005rb,Doring:2018cob,Kostelecky:2003cr,Katori:2006mz,Diaz:2010ft,Diaz:2011ia,Gninenko:2009ks,Gninenko:2009yf,Bai:2015ztj,Moss:2017pur,Bertuzzo:2018itn,Ballett:2018ynz,Fischer:2019fbw,Moulai:2019gpi,Dentler:2019dhz,deGouvea:2019qre,Datta:2020auq,Dutta:2020scq,Abdallah:2020biq,Abdullahi:2020nyr,Liao:2016reh,Carena:2017qhd,Abdallah:2020vgg, Abdallah:2024uby, Hammad:2021mpl, Gehrlein:2025zqq}. As these experiments continue to undergo further scrutiny, we make an attempt to present UV complete models for a pure axially coupled gauge boson with sub-GeV mass, which may be of interest in connection with them. There has also been general interest in low mass vector bosons coupling to leptons (see, for example, Refs.~\cite{DiLuzio:2025ojt, Hostert:2023tkg}).

We find that left-right symmetric universal seesaw models~\cite{Davidson:1987tr, Berezhiani:1985in, Babu:1989rb} provide a suitable venue for discussion of a UV complete model for a low mass axial vector gauge boson. We propose three classes of models and study some of their implications. They are based on the gauge group $SU(3)_c\times SU(2)_L\times SU(2)_R\times U(1)_X\times U(1)_a$, with $U(1)_a$ gauge boson having a purely axial coupling to fermions. Naturally, it does not contribute to the electric charge.  

Our explicit constructions exploit the hidden symmetries in the universal seesaw models, the minimal versions of which have a vector-like $U(1)_X$ gauge symmetry. They also have hidden global $B-L$ symmetry after symmetry breaking, since the Higgs doublets of the model are $B-L$ neutral. The first class of models we propose (called class A models) uses the axial counterpart of the $U(1)_X$ symmetry as a local symmetry, whereas the second class of models (called class B models) uses the gauged version of the axial $U(1)_{B-L}$ symmetry. The third model (called model C) uses the axial $B-3L_\tau$ as the new anomaly free axial gauge symmetry. We then present variations of these models which use type II seesaw for neutrino masses.  The $U(1)_a$ is family universal for models A and B. 

Among our results, we have found complete models for a low mass axial gauge boson and show how each of them can explain the small neutrino mass. In each of these cases, we add a vector like Dirac fermion, connected via the $U(1)_a$ symmetry to other particles in the models, which becomes the dark matter of the universe. The models also provide an embedding of the parity solution to the strong CP problem following~\cite{Babu:1989rb}. 

The gauge couplings and symmetry breaking scales in the model are theoretically unconstrained prior to symmetry breaking. However, in class A models, symmetry breaking links new physics to the $Z$ -boson properties and implies theoretical constraints on the gauge couplings depending on the new symmetry breaking scales, which we believe is a novel feature. In all these models, the axial gauge boson can have sub-GeV mass.

We then discuss some phenomenological implications of the model. 
In a follow-up paper~\cite{DKM}, we use dark matter production from kaon decay and a dark matter coupled scalar with two $U(1)_a$ gauge bosons (${\cal A}$) ($\phi{\cal A A}$) and two photons ($\phi \gamma\gamma$) couplings to resolve the MiniBooNE low energy excess in model C. 

The paper is organized as follows: in Sec.~\ref{sec:models}, we discuss the charge assignment of particles under the gauge symmetry for the three classes of models. In Sec.~\ref{sec:fermasses}, we discuss fermion masses, including those of neutrinos and strong CP solution in the model; in Sec.~\ref{sec:darkmatter}, we discuss how dark matter is explained in the model; Sec.~\ref{sec:symbreaking} is devoted to the discussion of gauge symmetry breaking and neutral gauge boson mixing. Sec.~\ref{sec:Aboson} is devoted to the discussion of the fermion couplings of the axial vector boson. In Sec.~\ref{sec:modelAuplimit}, we discuss how in the class A model, 
there
is an upper limit on the axial gauge boson coupling $g_a$. We 
discuss  phenomenological implications in Sec.~\ref{sec:phenomenology}.
We end with some comments on higher unification of the model in Sec.~\ref{sec:comments} and then conclude in Sec.~\ref{sec:conclusions}. In an appendix, we show that the vacuum expectation values (VEVs) of the singlet fields in the model are also real and therefore do not affect the strong CP solution.

\section{The Models}
\label{sec:models}
As already noted, the models use the gauge group $SU(3)_c\times SU(2)_L\times SU(2)_R\times U(1)_X\times U(1)_a$. The $U(1)_a$ part is the axial gauge symmetry with the associated gauge boson ${\cal A}_\mu$ having pure axial vector coupling with quarks and leptons, before symmetry breaking. If we choose
the associated gauge coupling to be $g_a \sim 10^{-2}- 10^{-4}$  and the symmetry breaking scale $\langle \eta \rangle\sim \mathcal{O}(\rm{TeV}) $, the axial gauge boson mass would be in the sub-GeV to 10 GeV range.

Below, we give three assignments for $U(1)_a$ charges, which lead to the anomaly-free models, all differing from each other in their Higgs content. We have assumed the $U(1)_a$ charges to be family universal, for the models A and B. As an illustrative example of a non-universal $U(1)_a$ scenario, we consider the case in which the gauged quantum number is $(B - 3L_{\tau,\mu})$ (Model C). In addition to its  distinctive phenomenology, this framework has the potential to explain the MiniBooNE low-energy excess~\cite{DKM}. 

\subsection{Model A}
This class of models uses the gauged version of the axial part of the $U(1)_X$ symmetry. The charge assignments of various particles under the gauge group are summarized in Table.~\ref{Tab:particle contentI}. 

\begin{table}
\centering
{\begin{tabular}{|c|c|c|c|c|c|c|}
		\hline \hline
		\bf Type &\bf Particle   & \bf{$\mathbf{SU(3)_C}$}  & \bf{$\mathbf{SU(2)_R}$} &\bf{$\mathbf{SU(2)_L}$} &\bf{$\mathbf{U(1)_{X}}$}&$\mathbf{U(1)_{a}}$\\
		\hline \hline
		{\bf Quarks} &$Q_{jL} = \left(
		\begin{array}{c}
			u_j\\
			d_j\\
		\end{array}
		\right)_L$  & \bf{3} &  \bf{1} &  \bf{2}  & 1/3& 1/3\\
		&$Q_{jR} = \left(
		\begin{array}{c}
			u_j\\
			d_j\\
		\end{array}
		\right)_R $  & \bf{3} &  \bf{2} &  \bf{1}  & 1/3&-1/3\\
		\hline
		{\bf Leptons} &$\psi_{jL} = \left(
		\begin{array}{c}
			\nu_j\\
			e_j\\
		\end{array}
		\right)_L$  & \bf{1} &  \bf{1} &  \bf{2}   & -1&-1\\
		&$\psi_{jR} = \left(
		\begin{array}{c}
			\nu_j\\
			e_j\\
		\end{array}
		\right)_R$  & \bf{1} &  \bf{2} &  \bf{1}   & -1&+1\\
		\hline \hline
		 &$U_R$ & \bf{3} &  \bf{1} &  \bf{1}  & 4/3&$4/3$\\
		{\textbf{BSM Fermions}} &$U_L$& \bf{3} &  \bf{1} &   \bf{1}  &$4/3$&$-4/3$\\
        &$D_R$ & \bf{3} &  \bf{1} &  \bf{1}  & $-2/3$&$-2/3$\\
		 &$D_L$& \bf{3} &  \bf{1} &   \bf{1}  &$-2/3$&$+2/3$\\
		&$E_R$  & \bf{1} &  \bf{1} &   \bf{1}  & $-2$&$-2$\\
		&$E_L$  & \bf{1} &  \bf{1} &   \bf{1}  & $-2$&$+2$\\
        &$N_{L, R}$ &   \bf{1} &  \bf{1} &   \bf{1}  & $0$&$0$\\
			\hline\hline
		{\textbf{Scalars}} &$\chi_L = \left(
		\begin{array}{c}
			\chi_L^{0}\\
			\chi_L^{-}\\
		\end{array}
		\right)$  & \bf{1} &  \bf{1} &  \bf{2}   & $ -1$ &$-1$\\
		&$\chi_R = \left(
		\begin{array}{c}
			\chi_R^{0}\\
			\chi_R^{-}\\
		\end{array}
		\right)$  & \bf{1} &  \bf{2} &  \bf{1}   & $-1$ & $+1$\\
		&$\eta_1$ & \bf{1} &  \bf{1} & \bf{1} &$0$&$+8/3$ \\
        &$\eta_2$ & \bf{1} &  \bf{1} & \bf{1} &$0$&$-4/3$ \\
        &$\eta_3$ & \bf{1} &  \bf{1} & \bf{1} &\bf{0}&\bf{-4} \\
		\hline \hline
	\end{tabular}}
	\caption{Summary of charge assignment under the gauge group of the particles for model A, where $j = 1,2,3$ labels the SM families, and we consider three generations of vector-like fermions $(U,~ D, ~N,~ E)$. $N$ is a spectator field in this case without any gauge quantum number. Note that the Higgs doublets $\chi_{L,R}$ share the $U(1)_a$ quantum number. As we will see below, this will have important implications for the gauge coupling $g_a$ .}
	\label{Tab:particle contentI}
\end{table}

\vskip0.2in
\noindent{\bf Anomaly cancellation:} It is easy to check that all gauge anomalies in the model cancel out. For example, denoting an anomaly by $A(U(1)_a G^2)$ with $G=SU(2)_{L,R}, U(1)_a, U(1)_X$ or $SU(3)_c$ or gravity, we have the following.

\begin{equation}
    \begin{aligned}
        A(U(1)_a SU(2)_L^2) &= 3\cdot 1/3-1=0;\quad 
  A(U(1)_a SU(2)_R^2)=-3\cdot 1/3+1=0\\ \nonumber
  A(U(1)^3_a) &=[2\cdot 3 \cdot \frac{1}{27}-2-3\cdot\frac{64}{27}+3\cdot \frac{8}{27}+8]_L+[2\cdot 3 \cdot \frac{1}{27}-2-3\cdot\frac{64}{27}+3\cdot \frac{8}{27}+8]_R=0 \\\nonumber
  A(U(1)_XU(1)^2_a) &=[2\cdot 3 \cdot \frac{1}{27}-2-3\cdot\frac{64}{27}+3\cdot \frac{8}{27}+8]_L-[2\cdot 3 \cdot \frac{1}{27}-2-3\cdot\frac{64}{27}+3\cdot \frac{8}{27}+8]_R=0 \\\nonumber
  A(U(1)_a SU(3)_c^2)&=[2\cdot 1/3-4/3+2/3]_L+[2\cdot 1/3-4/3+2/3]_R=0\\\nonumber
  A(U(1)_a [\text{gravity}]^2)&=[6\cdot 1/3-2-3\cdot 4/3+3\cdot 2/3+2]_L+[6\cdot 1/3-2-3\cdot 4/3+3\cdot 2/3+2]_R=0\\
   \end{aligned}
\end{equation}
\begin{equation}
    \begin{aligned}
 A(U(1)_a U(1)_X^2)=[6\cdot \frac{1}{3}\cdot\frac{1}{9}-2-3\cdot\frac{64}{27}+3\cdot\frac{8}{27}+8]_L+[6\cdot \frac{1}{3}\cdot\frac{1}{9}-2-3\cdot\frac{64}{27}+3\cdot\frac{8}{27}+8]_R=0
\end{aligned}
\end{equation}
 The subscripts $L$ and $R$ correspond to contributions from left and right chirality fermions to anomalies. We assume that under $L\leftrightarrow R$, $\chi_L\to \chi_R$ and the $\eta$ fields go to their complex conjugates.
Note that, in this case the SM singlet fermions $N_{L,R}$ do not have any $U(1)_a$ quantum number. The Yukawa couplings for model A are given by the Lagrangian:

\begin{equation}
    \begin{aligned}
        {\cal L}_Y=&~ h_u \bar{Q}_L{\chi}_LU_R+ h_d \bar{Q}_L\tilde{\chi}_LD_R+h_e \bar{\psi}_L\tilde{\chi}_LE_R+h_\nu \bar{\psi}_L{\chi}_LN_R~+\\ &f_u\bar{U}_L \eta^*_1 U_R+ f_d\bar{D}_L \eta^*_2 D_R+f_e\bar{E}_L\eta^*_3 E_R+M_D\bar{N_L}N_R~+ M_L{N}^T_L N_L\\ &+{L\leftrightarrow R}+h.c.
    \end{aligned}
\label{interaction}\end{equation}

where $\tilde\chi_{L,R}=i\tau_2\chi^*_{L,R}$ and the $N$ mass has both Dirac and Majorana type terms. The gauge symmetry is broken in stages: first the vev's of $\eta_{1,2,3}$ break the $U(1)_a$ symmetry. The rest of the symmetries are broken by the vacuum expectation values (vev) of $\chi_{L,R}$. 
We choose the following vacuum expectation values $\langle\chi^0_{L,R}\rangle=v_{L,R}/\sqrt{2}$; $\langle\eta_{1,2,3}\rangle= v_{\eta_{1,2,3}}/\sqrt{2}$. This breaks the gauge group down to $SU(3)_c\times U(1)_{em}$ with the electric charge formula given by 
\begin{eqnarray}
    Q=I_{3L}+I_{3R}+\frac{X}{2}
\label{charge}\end{eqnarray}
Note that $Q_a$, the generator of the $U(1)_a$ field does not contribute to the electric charge. This is to be expected since it is a purely axial charge.
The symmetry breaking also gives mass to the SM fermions via universal quark seesaw mechanism~\cite{Davidson:1987tr, Berezhiani:1985in}.

\subsection{Model $A'$, a  variation of model A without gauge singlet fermions:}
We note that in model A, the singlet fermions, $N$ do not have any gauge quantum number and do not play any role in anomaly cancellation. Their primary role, as we see below, is to give mass to neutrinos by a variation of the type~I seesaw mechanism~\cite{Minkowski:1977sc, Mohapatra:1979ia, Gell-Mann:1979vob, Yanagida:1980xy, Glashow:1979nm}.
However, if we used the type~II seesaw mechanism~\cite{Lazarides:1980nt, Schechter:1980gr, Mohapatra:1980yp} instead, there would be no need to have $N$s in the model to obtain neutrino masses. This motivates our model $A'$, where we do not have $N$-fermions but instead have triplet Higgs fields with quantum numbers $(1,3,1,+2, +2) \oplus (1,1,3,+2, -2)$,
denoted by symbols $\Delta_L\oplus \Delta_R$, to get small neutrino masses. Since the fermions contributing to the anomaly equations are the same as in model A, the anomaly cancellation is maintained.

\begin{table}[h]
\centering
{\begin{tabular}{|c|c|c|c|c|c|c|}
		\hline \hline
		\bf Type &\bf Particle   & \bf{$\mathbf{SU(3)_C}$}  & \bf{$\mathbf{SU(2)_R}$} &\bf{$\mathbf{SU(2)_L}$} &\bf{$\mathbf{U(1)_{X}}$}&$\mathbf{U(1)_{a}}$\\
		\hline \hline
		{\bf Quarks} &$Q_{jL} = \left(
		\begin{array}{c}
			u_j\\
			d_j\\
		\end{array}
		\right)_L$  & \bf{3} &  \bf{1} &  \bf{2}  & 1/3& 1/3\\
		&$Q_{jR} = \left(
		\begin{array}{c}
			u_j\\
			d_j\\
		\end{array}
		\right)_R $  & \bf{3} &  \bf{2} &  \bf{1}  & 1/3&-1/3\\
		\hline
		{\bf Leptons} &$\psi_{jL} = \left(
		\begin{array}{c}
			\nu_j\\
			e_j\\
		\end{array}
		\right)_L$  & \bf{1} &  \bf{1} &  \bf{2}   & -1&-1\\
		&$\psi_{jR} = \left(
		\begin{array}{c}
			\nu_j\\
			e_j\\
		\end{array}
		\right)_R$  & \bf{1} &  \bf{2} &  \bf{1}   & -1&+1\\
		\hline \hline
		 &$U_R$ & \bf{3} &  \bf{1} &  \bf{1}  & 4/3&$1/3$\\
		{\textbf{BSM Fermions}} &$U_L$& \bf{3} &  \bf{1} &   \bf{1}  &$4/3$&$-1/3$\\
        &$D_R$ & \bf{3} &  \bf{1} &  \bf{1}  & $-2/3$&$1/3$\\
		 &$D_L$& \bf{3} &  \bf{1} &   \bf{1}  &$-2/3$&$-1/3$\\
		&$E_R$  & \bf{1} &  \bf{1} &   \bf{1}  & $-2$&$-1$\\
		&$E_L$  & \bf{1} &  \bf{1} &   \bf{1}  & $-2$&$+1$\\
        &$N_{R}$ &   \bf{1} &  \bf{1} &   \bf{1}  & $0$&$-1$\\
        &$N_{L}$ &   \bf{1} &  \bf{1} &   \bf{1}  & $0$&$+1$\\
			\hline\hline
		{\textbf{Scalars}} &$\chi_L = \left(
		\begin{array}{c}
			\chi_L^{0}\\
			\chi_L^{-}\\
		\end{array}
		\right)$  & \bf{1} &  \bf{1} &  \bf{2}   & $-1$ &$0$\\
		&$\chi_R = \left(
		\begin{array}{c}
			\chi_R^{0}\\
			\chi_R^{-}\\
		\end{array}
		\right)$  & \bf{1} &  \bf{2} &  \bf{1}   & \bf{-1}&$0$\\
		&$\eta_1$ & \bf{1} &  \bf{1} & \bf{1} &$0$&$-2/3$ \\
        &$\eta_2$ & \bf{1} &  \bf{1} & \bf{1} &$0$&$-2$ \\
		\hline \hline
	\end{tabular}}
	\caption{The particle spectrum in model B, where $j = 1,2,3$ labels the SM families, that includes three generations of iso singlet vector-like fermions $(U,~ D, ~N,~ E)$. Note that, unlike model A, in this model, the standard model Higgs doublet $\chi_L$ does not have a non-zero $U(1)_a$ quantum number.}
	\label{Tab.particle contentII}
\end{table}

\subsection{Model B} This model uses axial $B-L$ as the gauge symmetry with the particle assignment under the gauge group given below in Table.~\ref{Tab.particle contentII}. The new feature of this model is that the heavy SM singlet  $N_{L,R}$ fermions have non-zero $U(1)_a$ quantum number. In this case, as in the case of model A, all gauge anomalies cancel as we explicitly show below, giving an alternative anomaly-free renormalizable model with a purely axial gauge boson. The axial gauge generator here is the axial $B-L$ symmetry. It does not contribute to electric charge, as in case A. 

\vskip0.2in
\noindent{\bf Anomaly cancellation:} We now check the anomaly cancellation in case B.

\begin{equation}
    \begin{aligned}
         A(U(1)_a SU(2)_L^2)=&~3\cdot 1/3-1=0\\
  A(U(1)_a SU(2)_R^2)=&~3\cdot 1/3-1=0\\
  A(U(1)^3_a)=&~[2\cdot 3 \cdot \frac{1}{27}-2-3\cdot\frac{1}{27}-3\cdot \frac{1}{27}+1+1]_L +[2\cdot 3 \cdot \frac{1}{27}-2-3\cdot\frac{1}{27}+3\cdot \frac{1}{27}+1+1]_R=0 \\
  A(U(1)_a SU(3)_c^2)=&~[2\cdot 1/3-1/3-1/3]_L-[2\cdot 1/3-1/3-1/3]_R=0\\
  A(U(1)_a [\text{gravity}]^2)=&~[6\cdot 1/3-2-3\cdot 1/3-3\cdot 1/3+1+1]_L-[6\cdot 1/3-2-3\cdot 1/3-3\cdot 1/3+1+1]_R=0\\
  A(U(1)_a U(1)^2_X)=&~[1/3\cdot 6\cdot 1/9-1\cdot 2\cdot 1-1/3\cdot 3\cdot16/9-1/3\cdot 3 \cdot 4/9+4\cdot 1]_L+ \\
  &~[1/3\cdot 6\cdot 1/9-1\cdot 2 \cdot 1-1/3\cdot 3 \cdot 16/9 -1/3\cdot 3 \cdot 4/9 +1 \cdot 4]_R =0\\
  A(U(1)^2_a U(1)_X)=&~[6 \cdot 1/3 \cdot 1/9 -1\cdot 2\cdot 1
  +4/3 \cdot 3 \cdot1/9 -2/3 \cdot 3 \cdot 1/9 -2\cdot 1]_L\\
 &-[6 \cdot 1/3 \cdot 1/9 -1\cdot 2\cdot 1
  +4/3 \cdot 3 \cdot1/9 -2/3 \cdot 3 \cdot 1/9 -2\cdot 1]_R=0\\
  A(U(1)^3_X)=&~0
    \end{aligned}
\end{equation}
The subscripts $L$ and $R$, as in model A, correspond to contributions from left and right-handed fermions to anomalies. We assume that under $L\leftrightarrow R$, $\chi_L\to \chi_R$ and the $\eta$ fields go to their complex conjugates.
 
The Yukawa couplings for model B are given by:
\begin{equation}
    \begin{aligned}
        {\cal L}_Y=&~ h_u \bar{Q}_L{\chi}_LU_R+ h_d \bar{Q}_L\tilde{\chi}_LD_R+h_e \bar{\psi}_L\tilde{\chi}_LE_R+h_e \bar{\psi}_L{\chi}_LN_R~+~{L\leftrightarrow R}+\\ & f_u\bar{U}_L \eta_1 U_R+f_d \bar{D}_L \eta^*_1 D_R+f_e\bar{E}_L\eta^*_2 E_R+f_N\bar{N}_L\eta^*_2 N_R~+ f_{Nm} N^T_R\eta^*_2 N_R+ f_{Nm} N_L^T\eta_2 N_L~+~h.c.
    \end{aligned}
\end{equation}

where $\tilde\chi=i\tau_2\chi^*$. This Lagrangian is left-right symmetric. The electric charge formula remains the same as in Eq.~\ref{charge}.

\subsection{Model C}
Model C has the axial gauge group $U(1)_{{B-3L_{\tau}}}$ where $B$ is the baryon number and $L_{\tau}$ is the tau generation lepton number. The quantum number assignments of particles under the gauge group are shown in Table.~\ref{Tab.particle contentIII}.

\begin{table}
\centering
{\begin{tabular}{|c|c|c|c|c|c|c|}
		\hline \hline
		\bf Type &\bf Particle   & \bf{$\mathbf{SU(3)_C}$}  & \bf{$\mathbf{SU(2)_R}$} &\bf{$\mathbf{SU(2)_L}$} &\bf{$\mathbf{U(1)_{X}}$}&$\mathbf{U(1)_{a}}$\\
		\hline \hline

		{\bf Quarks} &$Q_{jL} = \left(
		\begin{array}{c}
			u_j\\
			d_j\\
		\end{array}
		\right)_L (j=1,2,3)$ & \bf{3} &  \bf{1} &  \bf{2}  & 1/3& 1/3\\
		&$Q_{jR} = \left(
		\begin{array}{c}
			u_j\\
			d_j\\
		\end{array}
		\right)_R (j=1,2,3)$ & \bf{3} &  \bf{2} &  \bf{1}  & 1/3&-1/3\\
		\hline
		{\bf Leptons} &$\psi_{jL} = \left(
		\begin{array}{c}
			\nu_j\\
			e_j\\
		\end{array}
		\right)_L (j=1,2,3)$ & \bf{1} &  \bf{1} &  \bf{2}   & -1&$(0,0,-3)$\\
		&$\psi_{jR} = \left(
		\begin{array}{c}
			\nu_j\\
			e_j\\
		\end{array}
		\right)_R (j=1,2,3)$ & \bf{1} &  \bf{2} &  \bf{1}   & -1&$(0,0,+3)$\\
		\hline \hline
		 &$U_{j,R} (j=1,2,3)$ & \bf{3} &  \bf{1} &  \bf{1}  & 4/3&$1/3$\\
		{\textbf{BSM Fermions}} &$U_{jL}(j=1,2,3)$& \bf{3} &  \bf{1} &   \bf{1}  &$4/3$&$-1/3$\\
        &$D_{jR} (j=1,2,3)$ & \bf{3} &  \bf{1} &  \bf{1}  & $-2/3$&$+1/3$\\
		 &$D_{jL} (j=1,2,3)$& \bf{3} &  \bf{1} &   \bf{1}  &$-2/3$&$-1/3$\\
		&$E_{jR} (j=1,2,3)$  & \bf{1} &  \bf{1} &   \bf{1}  & $-2$&$(0,0,-3)$\\
		&$E_{jL} (j=1,2,3)$  & \bf{1} &  \bf{1} &   \bf{1}  & $-2$&$(0,0,+3)$\\
        &$N_{jR} (j=1,2,3)$ &   \bf{1} &  \bf{1} &   \bf{1}  & $0$&$(0,0,-3)$\\
        &$N_{jL} (j=1,2,3)$ &   \bf{1} &  \bf{1} &   \bf{1}  & $0$&$(0,0,+3)$\\
		
			\hline\hline
		{\textbf{Scalars}} &$\chi_L = \left(
		\begin{array}{c}
			\chi_L^{0}\\
			\chi_L^{-}\\
		\end{array}
		\right)$  & \bf{1} &  \bf{1} &  \bf{2}   & $ -1$ &$0$\\
		&$\chi_R = \left(
		\begin{array}{c}
			\chi_R^{0}\\
			\chi_R^{-}\\
		\end{array}
		\right)$  & \bf{1} &  \bf{2} &  \bf{1}   & $-1$ & $0$\\
		&$\eta_1$ & \bf{1} &  \bf{1} & \bf{1} &$0$&$+2/3$ \\
        &$\eta_2$ & \bf{1} &  \bf{1} & \bf{1} &\bf{0}&\bf{3} \\
        &$\eta_3$ & \bf{1} &  \bf{1} & \bf{1} &\bf{0}&\bf{6} \\
		\hline \hline
	\end{tabular}}
	\caption{Summary of charge assignment under the gsuge group of particles for model C (with axial $B-3L_\tau$), where $j = 1,2,3$ labels the fermion families. We consider three generations of vector-like fermions $(U,~ D, ~N,~ E)$. $N$ is a fermion field with only $U(1)_a$ gauge quantum number.}
	\label{Tab.particle contentIII}
\end{table}

\vskip0.2in
\noindent{\bf Anomaly cancellation:} It is easy to check that all gauge anomalies in the model cancel out once all three generations are included. We don't present the details here except for the least trivial cancellations of the $[U(1)_{B-3L_\tau}]^3$ and $[U(1)_{B-3L_\tau}]^2U(1)_X$. The anomalies are written for particles from top down, row by row, in Table.~\ref{Tab.particle contentIII}:
\begin{equation}
    \begin{aligned}
        A[(U(1)_{B-3L_\tau}^3)=&~6\times 3\times \frac{1}{27}+6\times 3\times \frac{1}{27}-2\times 27-2\times 27\\
        &-3\times 3 \times \frac{1}{27}-3\times 3 \times \frac{1}{27}-3\times 3 \times \frac{1}{27}-3\times 3 \times \frac{1}{27}+27+27+27+27=0\\
   A(U(1)_{B-3L_\tau}^2U(1)_X)=&~6\times \frac{1}{27}\times 3- 6\times \frac{1}{27}\times 3-2\times 9+2\times 9 \\
   &-\frac{4}{3}\times 9\times\frac{1}{9}+\frac{4}{3}\times 9\times\frac{1}{9}+\frac{2}{3}\times 9\times\frac{1}{9}-\frac{4}{3}\times 9\times\frac{1}{9}+2\times 9-2\times 9=0
    \end{aligned}
\end{equation}

The Yukawa couplings for model C are given by:
\begin{equation}
    \begin{aligned}
        {\cal L}_Y=&~ h_u \bar{Q}_L{\chi}_LU_R+ h_d \bar{Q}_L\tilde{\chi}_LD_R+h_{e, ii} \bar{\psi}_{iL}\tilde{\chi}_LE_{iR}+h_{\nu, 33} \bar{\psi}_{3L}{\chi}_LN_{3R}+h_{\nu,i3} \bar{\psi}_{iL}{\chi}_LN_{3R}\frac{\eta^*_2}{M}~+ f_u\bar{U}_L \eta^*_1 U_R\\
  &+f_d \bar{D}_L \eta^*_1 D_R+f_{33}\bar{E}_{3L}\eta_3 E_{3R}+\sum_{i=1,2}f_{3i}\bar{E}_{3L}\eta_2 E_{iR}
  +f_{N33}\bar{N}_{3L}\eta_3 N_{3R~}+f_{m3i}N^T_{3L}\eta_2 N_{iL}\\ &+\sum_{i=1,2}f_{N3i}\bar{N}_{3L}\eta_2 N_{iR} +f_{Nm33} N^T_{3R}\eta_3 N_{3R}+
  \sum_{i,j=1,2} [M^D_{N,ij}\bar{N}_{iL} N_{jR~}+ M^m_{N,ij} N^T_{iR} N_{jR}]\\
  &+\sum_{i,j=1,2}[ h_{ij} \bar{\psi}_{iL}{\chi}_LN_{jR}]~+~[{L\leftrightarrow R, \eta\leftrightarrow \eta^*}]
  ~+~h.c.
    \end{aligned}
\end{equation}

where $\tilde\chi=i\tau_2\chi^*$. After symmetry breaking, this Yukawa Lagrangian introduces mixings among different generations despite the starting symmetry picking out $L_\tau$ as part of the gauge group. The neutrino masses are realized as in Eq.~\ref{neutrinomass}. Note that the first two lepton generations in case C do not have any axial gauge interaction. Also incidentally, the $L_\tau$ in the axial gauge group could be $L_\mu$ as well with appropriate changes in the Yukawa Lagrangian.


\section{Fermion masses and strong CP} 
\label{sec:fermasses}
Both of the above models have parity (P) symmetry defined as follows:
\begin{eqnarray}
  Q_L\leftrightarrow Q_R; ~~\chi_L \leftrightarrow \chi_R;~~\eta_a \leftrightarrow \eta^*_a;~~{\cal A} \leftrightarrow {\cal - A}
\end{eqnarray}
The left- and right-handed Fermion Yukawa couplings are equal because of P-invariance, and the VEVs of $\chi_{L,R}$ are naturally real. In the appendix, we show that at the potential minimum, the $\eta_i$ VEVs are also real. It is then easy to see that, the resulting quark mass matrices have a real determinant. This leads to a solution of the strong CP problem as pointed out in ~\cite{Babu:1989rb}. Furthermore, the charged fermion masses are given by the seesaw formula~\cite{Davidson:1987tr}, 
\begin{equation}
    m_u\simeq \frac{h^2_u v_Lv_R}{f_Uv_\eta}
\end{equation}
and similarly for down quarks and charged leptons.
This makes the Yukawa couplings more natural than in SM. For example, the electron Yukawa coupling is $\geq 10^{-3}\sqrt{(f_Ev_\eta/v_R)}$ instead of its SM value of $\sim 10^{-6}$ and similarly other Yukawa couplings are larger, making it a less fine tuned model for fermion masses.

An important implication of universal quark seesaw for fermion masses in models A and B is that the value of $f_3v_\eta \leq v_R$ to get the right order of magnitude for the top quark mass ($m_t\sim v_L$).  If we choose $v_R$ in the few TeV range, this will limit the value of $v_\eta$ to be in the TeV range. In fact, if we want the Planck scale corrections not to destroy the strong CP solution, the $v_R \leq 100$ TeV~\cite{Berezhiani:1992pq}. This will have implications for the neutrino sector, as we will see below.

\subsection{Neutrino mass}

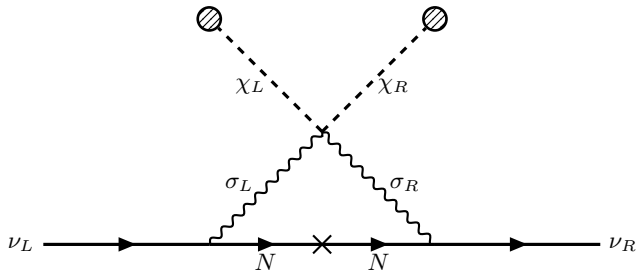
\begin{figure}
    \centering
    \begin{tikzpicture}
        \begin{feynman}

            \tikzset{
              bigblob/.style={
                draw,
                circle,
                minimum size=0.3cm,
                line width=1pt,
                pattern=north east lines,
                pattern color=black
              }
            }

            \vertex (i) at (-4, -1) {$\nu_L$};
            \vertex (f) at (4, -1) {$\nu_R$};
            \vertex (Ni) at (-1.5,-1);
            \vertex (x) at (0, -1);
            \vertex (Nf) at (1.5,-1);
            \vertex (o) at (0, 0.5);

            \vertex[bigblob] (o1) at (-1.5,2) {};
            \vertex[bigblob] (o2) at (1.5,2) {};

            \diagram*{
                (i) -- [fermion, very thick] (Ni)
                    -- [fermion, very thick, edge label'={$N$}] (x)
                    -- [fermion, very thick, edge label'={$N$}] (Nf)
                    -- [fermion, very thick] (f),

                (Ni) -- [boson, thick] (o)
                     -- [boson, thick] (Nf),

                (o) -- [scalar, very thick] (o1),
                (o) -- [scalar, very thick] (o2),
            };

            \node at (-1.1,-0.2) {$\sigma_L$};
            \node at (1.1,-0.2) {$\sigma_R$};
            \node at (-0.95,1.1) {$\chi_L$};
            \node at (0.95,1.1) {$\chi_R$};

            \draw[thick] ($(x)+(-0.12,-0.12)$) -- ($(x)+(0.12,0.12)$);
            \draw[thick] ($(x)+(-0.12,0.12)$) -- ($(x)+(0.12,-0.12)$);

        \end{feynman}
    \end{tikzpicture}
    \caption{One-loop Feynman diagram contributing to neutrino mass.}
    \label{fig:feynman_neumass}
\end{figure}

Neutrino masses are less trivial to understand in this class of models~\cite{Babu:2022ikf,Babu:2023dzz}. In our case, in model A, one can add arbitrary Majorana masses for $N_{L,R}$ as we have done in Eq.~\ref{interaction} because of which, smallness of neutrino masses is easier to understand in model A than in model B, where Majorana masses for $N_{L,R}$ are proportional to the value of $v_\eta$. 

The crucial observation here is that due to the different $U(1)_a$ quantum numbers of the $N_{L,R}$ fermions,in the two models, the understanding of smallness of neutrino masses is different. 

\subsubsection{ Model A} In this case, $N_{L,R}$ are neutral gauge singlets under all the gauge groups, one pair per family. At the renormalizable level, therefore, this allows arbitrary Majorana masses for $N$ together with arbitrary Dirac masses for $N_{L,R}$. The 
 structure of the $\nu-N$ mass matrix for one flavor is given in the basis $(\nu_L~\nu_R~ N_R ~ N_L)$ for model A by
\begin{eqnarray}
    M_{\nu-N}\simeq \left(\begin{array}{cccc}0 & 0 & h_\nu v_L & 0 \\0 & 0 & 0 & h_\nu v_R\\
    h_\nu v_L & 0 & M_R & \delta M\\0  & h_\nu v_R & \delta M & M_L\label{massmatrix}\end{array}\right)
\label{neutrinomass}\end{eqnarray}
Here we have used the two-component notation and denote left-and right- handed components of $\nu$ and $N$ by subscripts $L$ and $R$ respectively. For the case where $h_\nu \ll 1$, we can use the seesaw approximation to find the effective mass matrix for $(\nu_L~~\nu_R)$ and get,
\begin{eqnarray}
  M_{\nu_L\nu_R}~=~\frac{1}{M_LM_R-\delta M^2} \left(\begin{array}{cc} h^2_\nu v^2_L M_L & -h^2_\nu v_L v_R \delta M\\ -h^2_\nu   v_L v_R \delta M & h^2_\nu v^2_R M_R\end{array}\right)  
\end{eqnarray}
The smaller eigenvalues are given by (suppressing flavor indices)
\begin{eqnarray}
 m_{\nu_L, \nu_R}\simeq \frac{h^2_\nu v^2_{L,R}}{M_{R,L}}. 
 \end{eqnarray}

 Since the $M_{L,R}$ are free parameters, we can adjust them  to get right neutrino masses. For example, taking $h_\nu\sim 10^{-3}$ and taking $M_{L,R}$ to be $\sim 10^8-10^7$ GeV, we get the right order of magnitude for the known $\nu_1$ masses. The three $\nu_2$ eigenstates are the three sterile neutrinos with masses in the eV range. The $\nu_L-\nu_R$ mixing is very small for $\delta M\sim h_\nu v_L$ and vanishes if we set $\delta M=0$. 

 There is also a one loop contribution to both the Dirac and Majorana masses of $\nu_{L,R}$ coming from Fig.~\ref{fig:feynman_neumass}. For $h_\nu\sim 10^{-3}$, this contribution to neutrino masses is small. As far as the tree level Dirac mass is concerned, it is also small when the magnitude of $\delta M$ is small. 
 
\subsubsection{Model $A'$} In this case, the lepton Yukawa couplings are of the form
\begin{eqnarray}
  {\cal L}_Y~=~h_e \bar{\psi}_L{\chi}_LE_R+~{L\leftrightarrow R}+f\psi_L\psi_L\Delta_L+L\leftrightarrow R +h.c.
\end{eqnarray}
Once the $\Delta_{L,R}$ fields get vevs $\kappa_{L,R}$, both left- and right- handed neutrinos will pick up mass via the type II seesaw mechanism. Their masses are $m_{\nu_{L,R}}\simeq f\kappa_{L,R}$. The smallness of neutrino masses will come from the suppression of the triplet vevs ($\kappa_{L,R})$ as in the usual type II seesaw case. The $\Delta_R$ VEV could be chosen such that the RH neutrino masses could be heavier and decay to lighter fields during evolution of the universe and avoid any cosmological problems.
 
\subsubsection{ Model B} In this case, the Majorana and Dirac mass terms involving $N$'s require their being coupled to $\eta$ fields. The Dirac and Majorana masses are therefore proportional to $v_\eta$ and are limited by perturbativity to be less than or $\sim v_R$. 
 In the mass matrix of Eq.~\ref{massmatrix}, we have $M_{L,R}=fv_\eta$ and $\delta M=h_N v_\eta$. If we choose $v_R\sim v_\eta$ in the few TeV range, the only way to understand neutrino oscillation data would be to choose $h_\nu \leq 10^{-6}$. The neutrino Yukawas are, thus,  much smaller in model B than the charged fermion Yukawas, which are $\geq 10^{-3}$. 

\subsubsection{Model C} In this case, there is an intriguing possibility for generating light neutrino masses. Note that for this model to be anomaly free, we need only one pair of singlet fermions $N_{\tau~ L, R}$ and not three. If we keep only $N_\tau$, the neutrino sector of the model simplifies considerably. One can then rotate the $\nu_{i L}$ and $\nu_{i R}$ such that only one linear combination of $\nu_{i}$'s (call it $\nu_x\equiv \sum (\alpha_i \nu_i)$) appears in the mass matrix and the other two remain massless at the tree level. The $(\nu_{iL}, \nu_{iR}, N_3)$ mass matrix then looks like in the two component basis $(\nu_{xL}, \nu_{xR}, N_{3L} , N_{3R})$ as follows:
\begin{eqnarray}
    M_{\nu, N}~=~\left(\begin{array}{cccc} 0 & 0 & 0 & m_L\\0 & 0 & m_R & 0\\0 & m_R & M_L & M_2\\m_L & 0 & M_2 & M_R\end{array}\right)
\label{modelCneutrino}\end{eqnarray}
where $M\gg m$ and we have set $M_2=0$. $m_{L,R}=h_\nu v_{L,R}$. This leads to two light and two heavy neutrinos with masses for the light ones given by
$m_{\nu_{ L, R}}=\frac{h^2_\nu v^2_{L,R}}{M_{R,L}} $ at the tree level. Then at the one loop level, one more light neutrino pair picks up mass via a neutral Higgs boson exchange in the loop with a smaller mass compared to $\nu_x$. The third pair of neutrinos in this model remain massless, unless we include higher-dimensional operators to give them a tiny mass, which we can do. Since $v_R \leq v_\eta\sim $ 10 TeV or so, the masses of the left and right-handed Majorana neutrinos can be adjusted by making $h_\nu \sim 10^{-5}$.

The light right-handed neutrinos do not cause a problem with big bang nucleosynthesis if $v_R\sim v_\eta$ is in the multi-TeV range, which we chose above. Here, we have displayed the mechanism for understanding neutrino masses only schematically and have not carried out any detailed mixing angle and mass fitting with oscillation data.

\section{Dark matter}
\label{sec:darkmatter}
The simplest way to obtain dark matter in these models is to introduce an extra vector-like fermion $\zeta_{L,R}$ with an arbitrary  $U(1)_a$ charge $Q_a$. They are vector-like to avoid introducing anomalies. The $\zeta$-field then picks up an arbitrary Dirac mass $M$ of the form $\bar{\zeta}_L\zeta_R$. Its interaction with the ${\cal A}$ field will then have the purely vector-like form 
\begin{eqnarray}
    {\cal L}_a \simeq ig_\zeta{\cal A}_\mu (\bar{\zeta}_{L}\gamma^{\mu}\zeta_{L}+\bar{\zeta}_{R}\gamma^{\mu}\zeta_{R}
    )
\end{eqnarray}
The thermal relic density of DM $\zeta$ can then be obtained via the annihilation channel $\zeta\bar{\zeta}\to  f \bar{f}$, $\zeta\bar{\zeta}\to {\cal A A}$ etc. by a suitable adjustment of the gauge coupling, $g_\zeta$, the strength of dark matter coupling to ${\cal A}$. 
For the thermal freeze-out scenario, to obtain the observed relic density, one must have~\cite{Mohapatra:2019ysk}
\begin{equation}
    g_{\zeta} \sim 0.016 \sqrt{m_\zeta ~{\rm in~ GeV}}
\end{equation}
Thus clearly, the dark matter particle must have O(1) $U(1)_a$ charge, $Q_a$ if the gauge coupling $g_a\leq 10^{-2}$ since $g_\zeta=Q_ag_a$. We will see in ~\cite{DKM}  that a large $Q_a$ for dark matter is also required to explain the MiniBooNE low energy excess in model C using dark matter scattering.

The origin of dark matter can also be freeze-in~\cite{Hall:2009bx,Chu:2011be} in this model. For example, with $g_a\sim10^{-6}$, it is possible to explain the DM abundance with $m_{\zeta}>m_{\cal A}/2$ as shown in a similar setup in the context of $U(1)_{B-L}$ vector gauge boson model~\cite{Mohapatra:2019ysk,Heeba:2019jho}. In this scenario, the gauge boson $A'$ is in thermal equilibrium, but not the dark matter. It is also possible to find the correct abundance in the context where both $A$ and DM are not in thermal equilibrium~\cite{Mohapatra:2019ysk}.

\section{Symmetry breaking and interactions} 
\label{sec:symbreaking}
In this section, we will discuss the symmetry breaking, and the resulting gauge-boson masses and gauge interactions. We will do this for models A, B and C.

The masses of the charged gauge bosons $W^{\pm}_{L,R}$ are given by
$M_{W^{\pm}_{L,R}}=\frac{gv_{L,R}}{2}$. There are four neutral gauge bosons $W_{3L, 3R}$, $B$  and ${\cal A}$ and the first three mix among themselves as in the usual left-right models with doublet Higgs fields, and in model A, they all mix among themselves. We write down the $4\times 4$ mass matrix for model A and consider its eigenvectors and eigenvalues.

In case A, the axial vector gauge boson $\cal A$ mixes with the left-right gauge bosons, $W_{3 L,R}$ and $B$ leading to the following neutral gauge boson mass matrix in the basis $(W_{3L}, W_{3R}, B , {\cal A})$:
\begin{eqnarray}
    {\cal M}^2~=~\frac{1}{4}\left(\begin{array}{cccc}g^2v^2_L & 0 &-gg'v^2_L & -gg_a v^2_L\\0 & g^2v^2_R & -gg'v^2_R & gg_av^2_R\\ -gg'v^2_L &  -gg'v^2_R & g^{'2}(v^2_L+v^2_R) & g'g_a ((v^2_L - v^2_R)\\ -gg_a v^2_L & gg_a v^2_R & g'g_a ((v^2_L - v^2_R) & g^2_a(v^2_L - v^2_R+a v^2_\eta) \end{array}\right)
\label{gaugebosonmass}\end{eqnarray}
where we have taken $g_a/2$ as the axial $U(1)_a$ gauge coupling. We have $av^2_\eta=64/9 v^2_{\eta_1}+16/9v^2_{\eta_2}+16v^2_{\eta_3}$. If we take all $v_\eta$'s to be equal, we find that $av^2_\eta=\frac{224}{9}v^2_\eta$.

We will use the short-hand notations, $s_w, c_w = \sin\theta_W, \cos\theta_W$, to denote the Weinberg angles. The massless eigenvector of this matrix (the photon) is given by
$(s_w, s_w, \sqrt{c^2_w-s^2_w}, 0)^T$, where $e/g=s_w$. After decoupling the massless mode, the eigenvalues and eigenvectors of the resulting $3\times 3$ matrix correspond to the masses and eigenstates of the three physical massive gauge bosons $(Z, Z_R, {\cal A'})$. We simply implement this using the following $4\times 4$ orthogonal matrix $O$~\cite{Jana:2024icm}:

\begin{eqnarray}
  O~=~\left(\begin{array}{cccc}s_w &s_w & \sqrt{c_{2w}}&0\\c_w &-s^2_w/c_w& -\frac{s_w}{c_w}\sqrt{c_{2w}}&0\\0 &\sqrt{c_{2w}}/c_w& -s_w/c_w &0\\0&0&0&1 
      \end{array}\right)
\end{eqnarray}

where $s_w=e/g$; $\sqrt{c_{2w}}=e/g'$.
We calculate $O^T{\cal M}^2 O$ and "take out" the massless eigenstate.
The remaining $3\times 3$ matrix for massive states is then given by:
\begin{eqnarray}
    \tilde{\cal {M}}^2~=~\frac{1}{4}\left(\begin{array}{ccc} A_{LL} v^2_L &   A_{LR} v^2_L& A_{La}v^2_L\\  A_{LR} v^2_L & A_{RR}v^2_R& A_{Ra}v^2_R+A'_{La} v^2_L\\
    A_{La} v^2_L &A_{Ra}v^2_R+A'_{La} v^2_L& A_{aa}\tilde{v}^2_\eta\end{array}\right)
\end{eqnarray}
where 
\begin{equation}
    \begin{aligned}
        A_{LL}=&~\frac{e^2}{s^2_wc^2_w};~
        A_{LR}=\frac{eg'}{c^2_w};\\
        A_{RR}=&~2egs_w/c^2_w+g^2{c_{2w}}/c^2_w
        +e^2s^2_w/c_{2w}c^2_w+e^2s^2_w/(c_{2w}c^2_w)(v^2_L/v^2_R);\\
        A_{La}=&~\frac{-eg_a }{c_ws_w};~
        A_{Ra}=-gg_a\sqrt{c_{2w}}/c_w-g'g_as_w/c_w\\
        A'_{La}=&-g'g_as_w/c_w;~~A_{aa}=g^2_a \tilde{v}^2_\eta;\\
        \tilde{v}^2_\eta=&~(v^2_L+v^2_R+\frac{224}{9} v^2_\eta)
    \end{aligned}
\end{equation}

 The eigenvectors and eigenvalues of this matrix give the physical states $Z, Z_R, {\cal {A'}}$ and their masses. Note that the $Z$ boson that can be identified with the SM $Z$ boson has the mass that coincides with that in the SM up to corrections of order $(v^2_L/v^2_R)$ for small $g_a\ll g,g'$. In this basis, it is easier to calculate the interaction of the $\cal A$ boson with quarks.

\section{ ${\cal A}$ boson couplings}

\label{sec:Aboson}
 Before symmetry breaking, the coupling of the axial gauge boson ${\cal A}$ to SM fermions is purely axial and given by
 \begin{eqnarray}
    {\cal L}_{\cal A}~=~-ig_a{\cal A}_\mu\sum_{f}Q_f\bar{f}\gamma^\mu \gamma_5 f   
 \end{eqnarray}
After symmetry breaking, the fermions mix among themselves, and the gauge boson ${\cal A}$ also mixes with other gauge bosons in model A. We will call the final physical gauge boson ${\cal A'}$ in model A and ${\cal A}$ in models B and C. This changes the fermion couplings of the physical axial gauge boson, ${\cal A'}$ (In subsequent pages, we will use ${\cal A}$ since we will consider the phenomenology of models B and C only). 

First note that due to the universal nature of the axial quantum number of  SM fermions, the CKM mixing does not affect the ${\cal A}$ coupling to fermions. There is, however, an effect due to the heavy and light quark mixings since the left and right heavy-light mixings of the fermions are chirality sensitive. Typically, the left mixings are of the order $\theta_L\simeq h v_L/fv_{\eta}$, while the right mixings are of the order $\theta_R\simeq h v_R/fv_{\eta}$. Thus, the right-handed heavy-light mixings $\theta_R$ are larger. If $v_\eta\sim 10 v_R$, the mixing effects are small for the first and second generation quarks and leptons. It will be larger for the third generation quarks. If we ignore the heavy-light mixing angles,  the physical ${\cal A'}$ couplings to the fermions will be dominantly axial. However, when we consider the decay of pion and $\eta$ mediated by ${\cal A}$, heavy-light mixing is the only one that counts, since the dominant contribution is an isospin singlet and a $SU(3)$ singlet. We will keep the dominant heavy-light mixing effects there.

Finally, in model A, the mixing of various gauge bosons with ${\cal A}$ will introduce some changes to this coupling as well, and they are shown below. The ${\cal A'}$ coupling to fermions in this model is therefore given by:
\begin{eqnarray}
    {\cal L}_{\cal A'}~=~-ig_a{\cal A'}_\mu\sum_{f}[\bar{f}\gamma^\mu(\gamma_5Q_f + \epsilon_Z I_{Z}+\epsilon_{Z_R} I_{Z_R})f
    +\epsilon_{fR} \bar{f}_R\gamma^\mu f_R]+h.c.
\end{eqnarray}
where $\epsilon_{Z,Z_R}$ are the small mixing coefficients between $U(1)_a$ gauge boson with $(Z,Z_R)$ gauge bosons and $\epsilon_{fR}\simeq \frac{h_f v_R}{f_\eta v_\eta}$ is the heavy-light fermion mixing effect. The dominant $Z$ mixing effect, $\epsilon_Z$, is given by:
\begin{eqnarray}
 \epsilon_Z=\frac{g_a}{e}s_wc_w.  
\end{eqnarray}
 Note that in this model, there will be a small admixture of vector currents to ${\cal A}$ through the admixture terms $\epsilon_Z$ and the heavy-light fermion mixing effects. This effect is not present in models B and C, whose phenomenology we discuss below
\footnote{For partially axial $U(1)$ gauge models with gauge boson coupling to both vector and axial currents, see ~\cite{Kahn:2016vjr}; \cite{ Ismail:2016tod}. In contrast, in our model, the new gauge boson ${\cal A}$ couples to pure axial currents prior to symmetry breaking.}.

{\bf Model B:} The neutral gauge boson mass matrix for this case is the same matrix as in Eq.~\ref{gaugebosonmass} except that the top three entries in the 4th column and the first three entries in the 4th row vanish, making ${\cal A}$ decouple from the other gauge bosons. The $3\times 3$ mass matrix involving $W_{3 L,R}$ and $B$ is the same as in minimal left-right models with doublet Higgs fields and its eigenstates and eigenvalues can be found in the literature (see, for example, ~\cite{Mohapatra:1998rq, Jana:2024icm}). In this case, the fermion coupling of $\cal A$, can be read from the gauge quantum numbers $Q_a$ of the various fermions, $f$. Ignoring heavy-light quark mixings, this interaction looks as follows: 
\begin{eqnarray}
    {\cal L}_{\cal A}~=~-ig_a{\cal A}_\mu\sum_{f}Q_f\bar{f}\gamma^\mu\gamma_5f
\end{eqnarray}

The ${\cal A}$ gauge boson couplings in the $A'$ model are similar to the $A$ model, while the couplings in the $C$ model are similar to the $B$ model.

\section{Upper limit on $g_a$ in model A}
\label{sec:modelAuplimit}
A novel implication of our model is that due to the SM Higgs doublet sharing the $U(1)_a$ gauge quantum number, the SM $Z$-boson mass and its couplings are affected by the values of $g_a$ as well as the symmetry-breaking VEV $v_\eta$. As we increase $g_a$ from zero, there is a point in the $g_a$ axis where the lightest eigenvalue crosses $Z$  and $Z_R$, changing the $Z$ eigenvalue and the corresponding eigenvector, drastically modifying the properties of the $Z$ boson. We call this the ``level crossing" point.  The observed properties of the $Z$-boson therefore lead to an upper limit on $g_a$. This happens for model A and $A'$ and arises because the mixings between $W_{3L, 3R}$, $B$ and ${\cal A}$ are induced by the VEVs of the doublet Higgs fields $\chi_{L,R}$. Models B and C do not have this property. We believe that this is a general result for any theory in which the SM Higgs doublet shares the gauge quantum numbers of another $U(1)$ group.

\begin{figure}
    \centering
    \subfloat[]{%
        \includegraphics[width=0.48\linewidth]{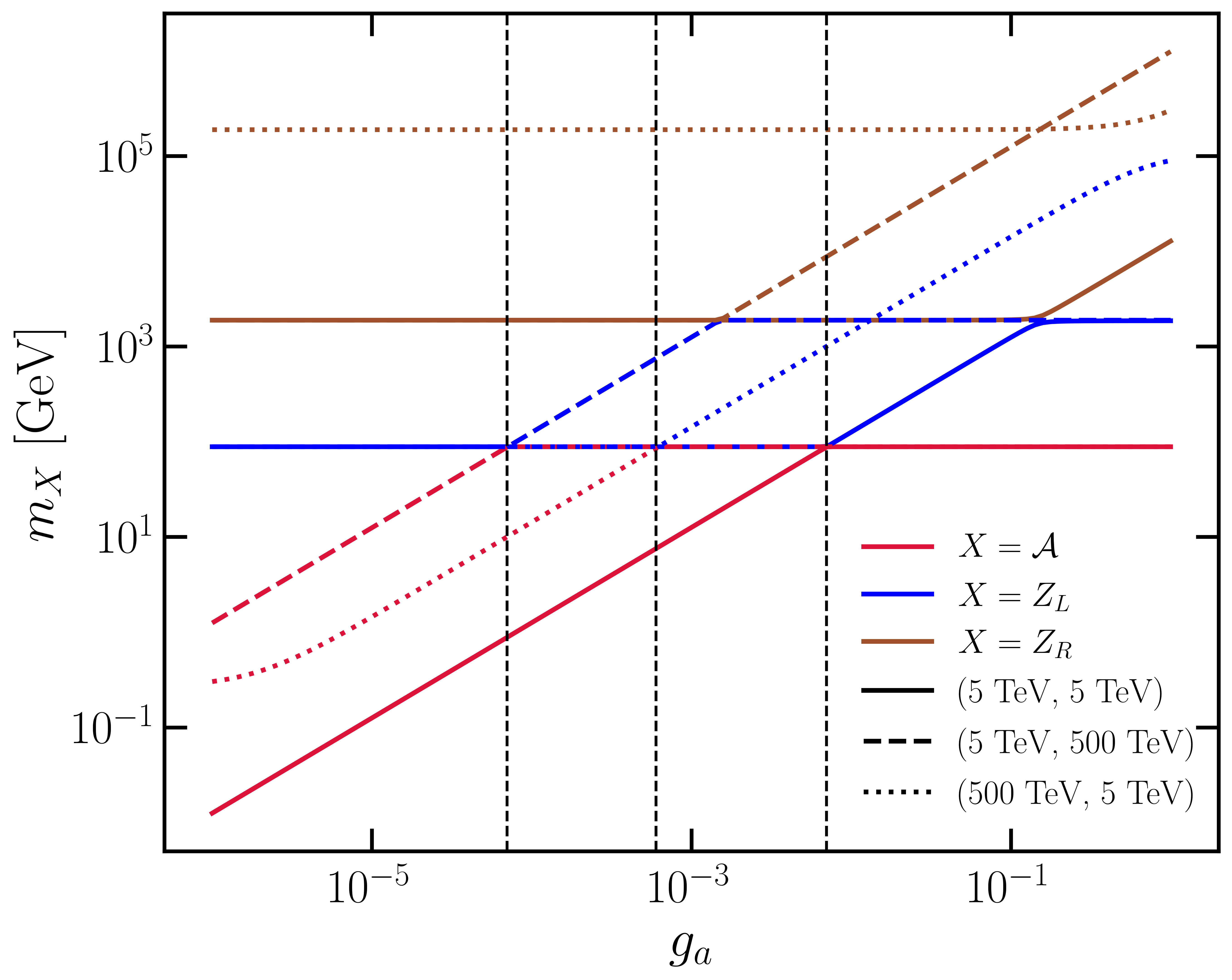}
        \label{fig:masses}
    }
    \hfill
    \subfloat[]{%
        \includegraphics[width=0.48\linewidth]{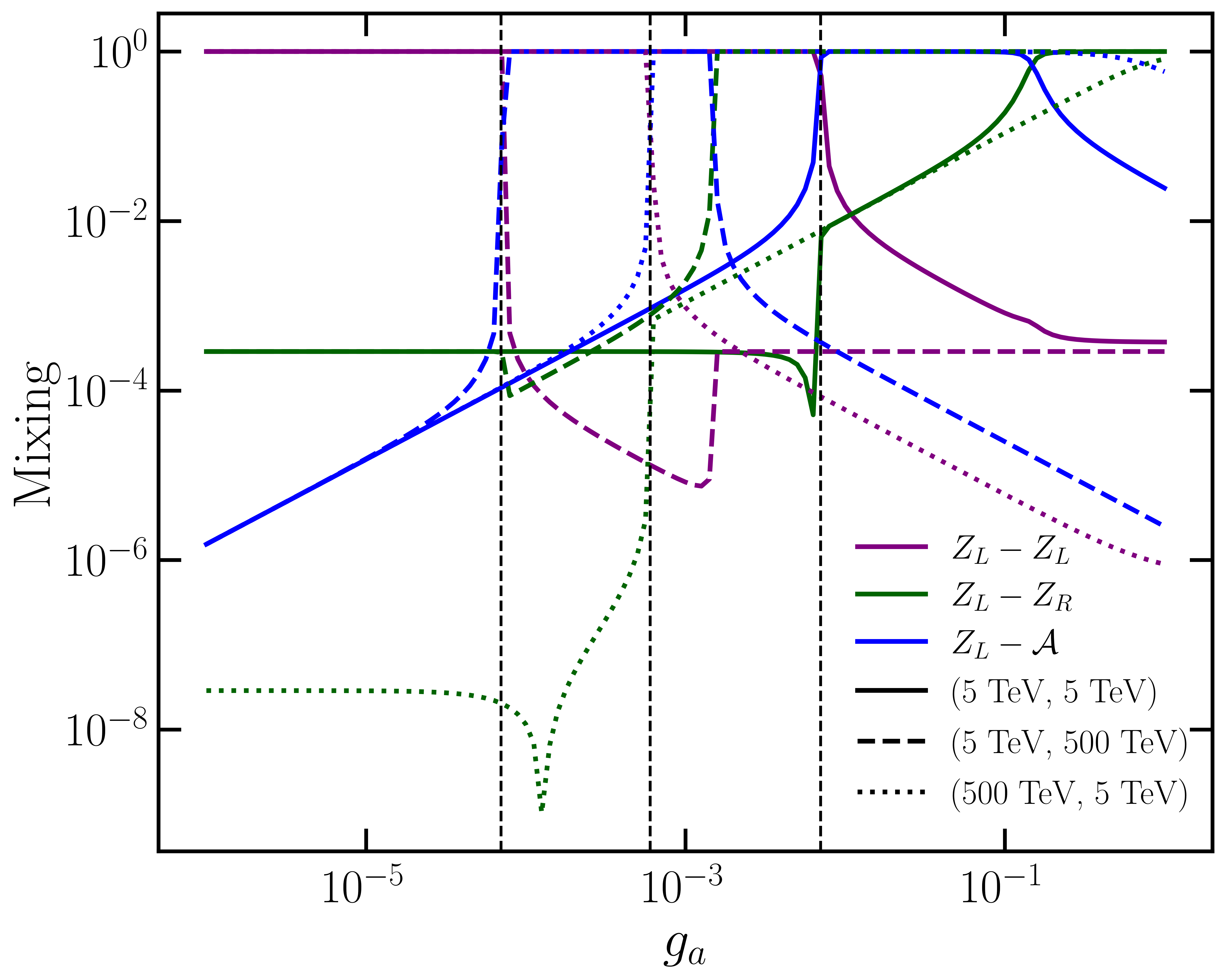}
        \label{fig:mixing}
    }
    \caption{(a) Variation of eigenvalues of the neutral gauge boson mass matrix as $g_a$ is increased. Note the level crossings for different values of $g_a$. The tuples ($v_R$, $v_\eta$) denote the choice of VEVs. The vertical lines denote the values of $g_a$ where the level crossing occurs giving the upper limits on $g_a$. This feature does not apply to model B and C where ${\cal A}$ does not mix with $Z$ and $Z_R$. (b) Variation of mixing angles of the $Z$ as $g_a$ is increased. The vertical lines denote the values of $g_a$ where the mixing of $Z$ changes drastically, giving the upper limits on $g_a$.}
    \label{fig:eigenvalues}
\end{figure}

To see intuitively how this limit arises, let us consider a simpler model by taking a $2\times 2$ matrix of the form:
\begin{eqnarray}
   \tilde{\cal {M}}^2_{eff}~=~\left(\begin{array}{cc} A_{LL} v^2_L &  A_{La}v^2_L\\   
    A_{La} v^2_L & A_{aa}\tilde{v}^2_\eta\end{array}\right) 
\end{eqnarray}
where we symbolically denote the dependence on $g_a$ as $A_{La}=cg_a$ and $A_{aa}=dg^2_a$. This mimics the $2\times 2$ sub-matrix of Eq.~\ref{gaugebosonmass}. We see from an analysis of this matrix that as $g_a\to 0$, the larger eigenvalue is $A_{LL}v^2_L$ and as $g_a\to \infty$, the smaller eigenvalue is also the same as $A_{LL}v^2_L$, whereas the larger eigenvalue exceeds the smaller one. There is a level crossing for some value of $g_a$. This is also reflected in our calculation of the $3\times 3$ case, as well as shown in Fig.~\ref{fig:eigenvalues}. From this, we find a maximum value of $g_a$ for which the lower mass eigenstate remains very close to $A_{LL}v^2_L$, which is its SM value for $Z$. This gives the upper limit on $g_a$ for model A and $A'$.

In Fig.~\ref{fig:masses}, we show what happens to the neutral vector boson mass eigenvalues in the $3\times 3$ case, as we increase $g_a$ from zero to very large values. This is the result of our calculation, whereas the $2\times 2$ case is done analytically. We see that for $g_a$ above certain values denoted by vertical lines in Fig.~\ref{fig:masses}, the value of $M_Z$ changes drastically. This gives the upper limit on $g_a$ for the theory. This upper limit depends on the choice of $v_R$ and $v_\eta$ as shown in Fig.~\ref{fig:masses}. The limits on $g_a$ for various values of $v_{R}$ and $v_\eta$ are given in Table.~\ref{tab:upplim_ga}.

\begin{table}[h]
\centering
    \begin{tabular}{|c|c|c|}
        \hline
        $g_a^{\text{max}}$ & $v_R~[{\rm TeV}]$ & $v_\eta~[{\rm TeV}]$\\ 
        \hline\hline
        $7\times 10^{-3}$ & $5~{\rm TeV}$ & $5 ~{\rm TeV}$\\
        \hline
        $6.2\times 10^{-4}$ & $500~{\rm TeV}$ & $5 ~{\rm TeV}$\\
        \hline
        $7\times 10^{-5}$ & $5~{\rm TeV}$ & $500 ~{\rm TeV}$\\
        \hline
   \end{tabular}
   
    \caption{Upper limits on $g_a$ from neutral gauge boson mixing in model A}
    \label{tab:upplim_ga}
\end{table}

The same happens also for mixing of the $Z$-boson with 
$Z_R$ and ${\cal A}$ (see Fig.~\ref{fig:mixing}). The values of $g_a$ when the mixing angles change drastically are shown by vertical lines in Fig.~\ref{fig:mixing}. The transition values of $g_a$ coincide for both masses and mixing angles.

\section{Phenomenology: low energy constraints} 
\label{sec:phenomenology}

In this section, we analyze the phenomenological bounds on the universal $U(1)_a$ model, for Case B, for MeV-GeV scale ${{\cal{A}}}$s. Since for model A, due to the mixing of different gauge bosons (Fig.~\ref{fig:eigenvalues}), the connection between $g_a$ and $M_{{\cal{A}}'}$ is more involved, and the parameter-space plot would involve a complicated function of different VEVs and couplings. On the other hand, in Model B (and also Model C), the parameter space can be fully characterized in terms of $g_{a}$ and $m_{\mathcal{A}}$. Moreover, in these models, $g_{a}$  is not subject to an upper bound.

\subsection{Universal $U(1)_a$} We will consider the case where the $U(1)_a$ charge assignments to fermions are family universal. e.g., model B. 

\medskip

\noindent\textbf{FCNC constraints} In universal $U(1)_a$ models, after symmetry breaking, quark seesaw introduces small departures from universality of the ${\cal A}$ couplings to flavors. The extent of non-universality before CKM rotation is given by $\sim \frac{m_q}{v_{wk}}$ in the coupling of ${\cal A}$ to quark currents for the choice $fv_\eta=v_R$. The right chirality interaction gives the dominant contribution, since the mixing of the light with the heavy quarks is chirality dependent and is given by $\theta_{L,R}\sim \frac{h_q v_{L,R}}{f_qv_\eta}$ and $v_L \ll v_R$. Using this mixing, one can estimate the different FCNC effects. For example, the $\Delta C=2$ Lagrangian in the model is given by 
\begin{eqnarray}
   {\cal L}_{\Delta C}= C_D (\bar{c}_R\gamma_\mu  {c}_R)^2
\end{eqnarray}
where $C_D\simeq (\frac{m_q}{v_{wk}}V_{cu})^2/ v^2_\eta$. Present data for $D-\bar{D}$ mixing then leads to $v_\eta\geq 2$ TeV.  This implies that for $g_a\sim 10^{-2}-10^{-4}$, the ${\cal A}$ mass in the sub-GeV to GeV range. One can look at other FCNC operators e. g. $\Delta S=2$ $K-\bar{K}$ 
operator. The $K-\bar{K}$ mixing from the corresponding $\Delta S=2$ operator implies a  weaker lower bound on $v_\eta$.

\medskip

We depict the allowed parameter space and constraints for the case where $\mathcal{A}$ rapidly decays to dark matter in Fig.~\ref{fig:bounds_univ}, and for the case where it doesn't decay to dark matter in Fig.~\ref{fig:bounds_univ_vis}. We will refer to the former scenario as an \textit{invisible} $\mathcal{A}$ (commonly when  $m_{\zeta} < m_{\mathcal{A}}/2$ and $g_{\zeta} \sim \mathcal{O}(1)$), and the latter a \textit{visible} $\mathcal{A}$ (such as when $m_{\zeta} > m_{\mathcal{A}}/2$, or when $m_{\zeta} < m_{\mathcal{A}}/2$ but $g_{\zeta} \ll g_a$). We will first discuss the constraints that apply in both scenarios, and then separately for the invisible and visible scenarios.

\medskip

\subsubsection*{Invisible and Visible bounds}
\begin{figure}
    \centering
    \subfloat[]{%
        \includegraphics[width=0.49\textwidth]{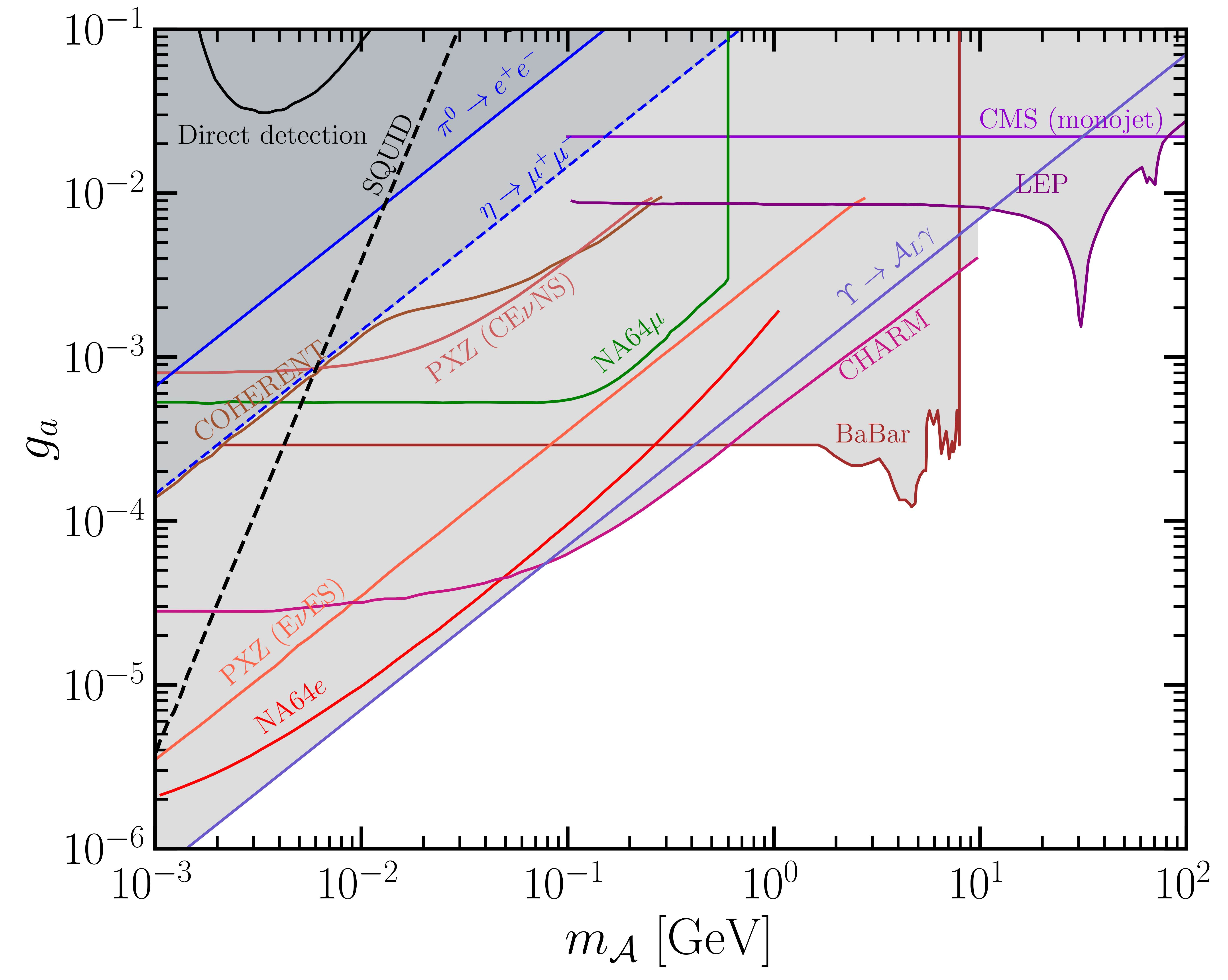}
        \label{fig:bounds_univ}
    }
    \hfill
    \subfloat[]{%
        \includegraphics[width=0.49\textwidth]{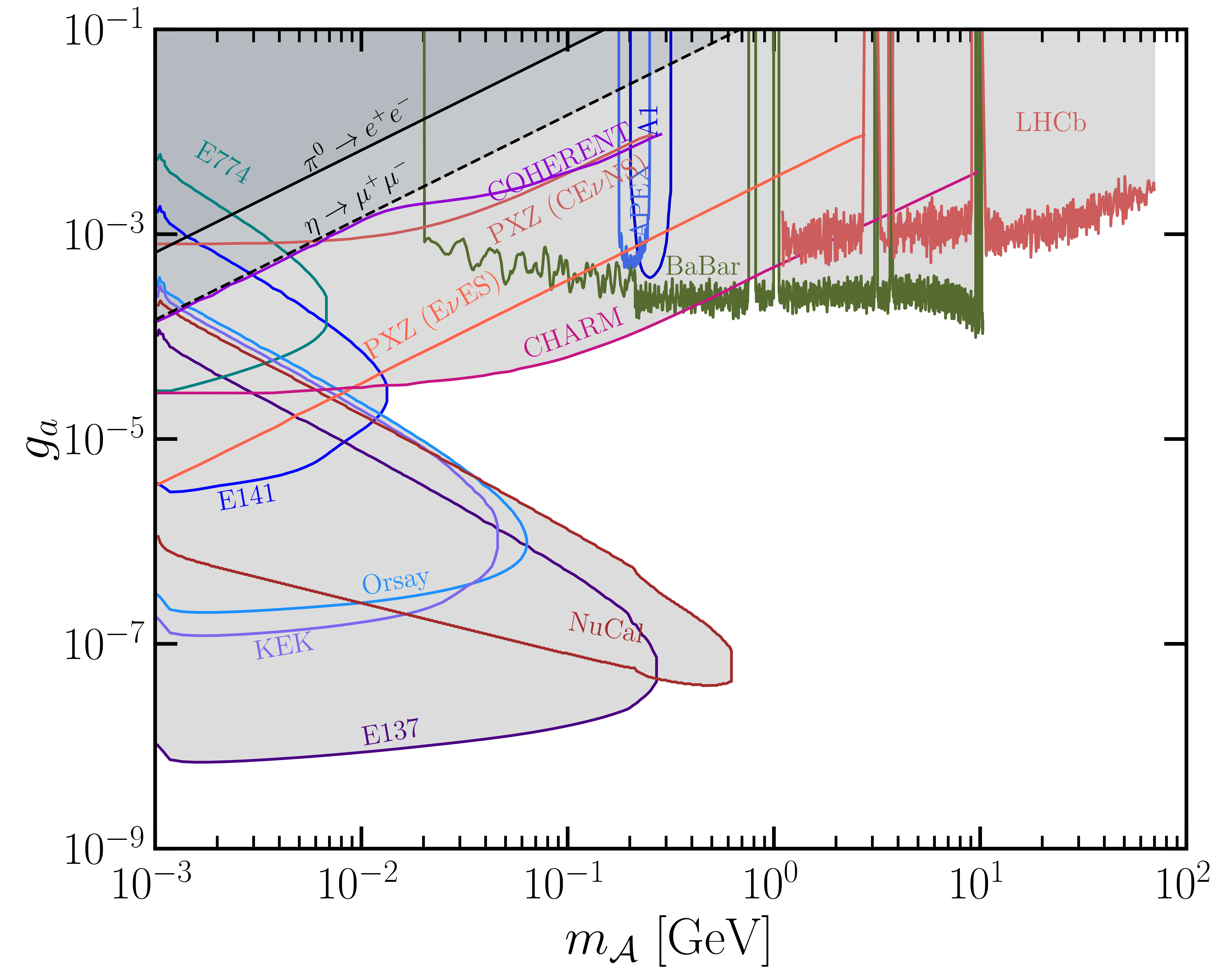}
        \label{fig:bounds_univ_vis}
    }
    \caption{Bounds on (a) universal invisibly decaying and (b) universal visibly decaying $B-L$ axial gauge boson (model B).}
    \label{fig:bounds_univ_combined}
\end{figure}

\noindent {\textbf{Solar Neutrino Scattering:}} The sensitivities of the latest generation dark matter detectors, such as PandaX-4T~\cite{PandaX-4T:2021bab}, XENONnT~\cite{XENON:2023cxc}, and LUX-ZEPLIN (LZ)~\cite{LZ:2022lsv}, have started to probe cross sections that are so low that they can probe the neutrino fog~\cite{OHare:2021utq}. Recent studies have looked at Coherent Elastic Neutrino-Nucleus (CE$\nu$NS) as well as Elastic Neutrino-Electron (E$\nu$ES) Scattering of solar neutrinos (all three flavors) at these detectors, and therefore placed limits on new neutrino interactions through new neutrino, electron, up, and down philic mediators~\cite{DeRomeri:2024iaw}. We therefore present the leading limits from a combined analysis of CE$\nu$NS and E$\nu$ES (denoted as PXZ in Fig.~\ref{fig:bounds_univ} and~\ref{fig:bounds_univ_vis} ), that probe $1~\text{MeV} < m_{\mathcal{A}} < 1~\text{GeV}$. Due to the axial nature of $\mathcal{A}$, the CE$\nu$NS cross section is suppressed~\cite{Cirelli:2013ufw, DelNobile:2021wmp}, accounting for the fact that only the spin-$1/2$ and $3/2$ isotopes contribute to the cross section, as well as the lack of enhancement from the atomic number and mass. Due to this reason, the E$\nu$ES dominates.

\medskip

\noindent \textbf{Accelerator Neutrino Scattering.} Neutrino experiments that operate at proton-on-target facilities contain plenty of neutrinos that are produced from charged meson decays, which undergo CE$\nu$NS as well as E$\nu$ES at the detector. We depict the bounds from COHERENT~\cite{AtzoriCorona:2022moj, DeRomeri:2022twg}, which mainly look for muon-neutrino CE$\nu$NS, as well as CHARM~\cite{Bauer:2018onh}, which probes muon-neutrino E$\nu$ES, in Fig.~\ref{fig:bounds_univ} and ~\ref{fig:bounds_univ_vis}. Since CHARM operates at an energy higher than that of COHERENT, the former can probe $m_\mathcal{A}$ up to $\mathcal{O}(10~\text{GeV})$.

\medskip


\begin{figure}
    \centering
    \subfloat[]{%
        \includegraphics[width=0.49\textwidth]{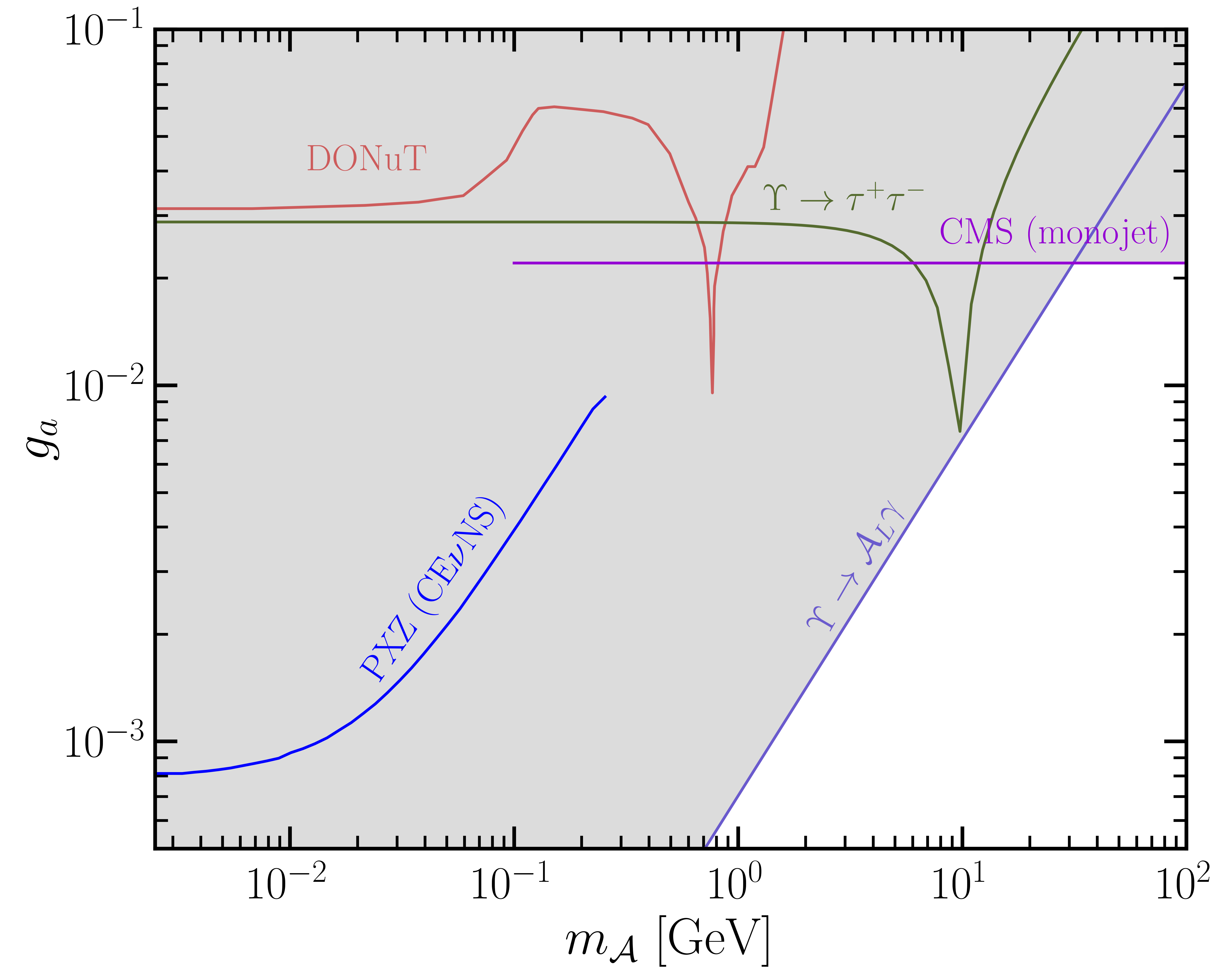}
        \label{fig:bounds_13}
    }
    \hfill
    \subfloat[]{%
        \includegraphics[width=0.49\textwidth]{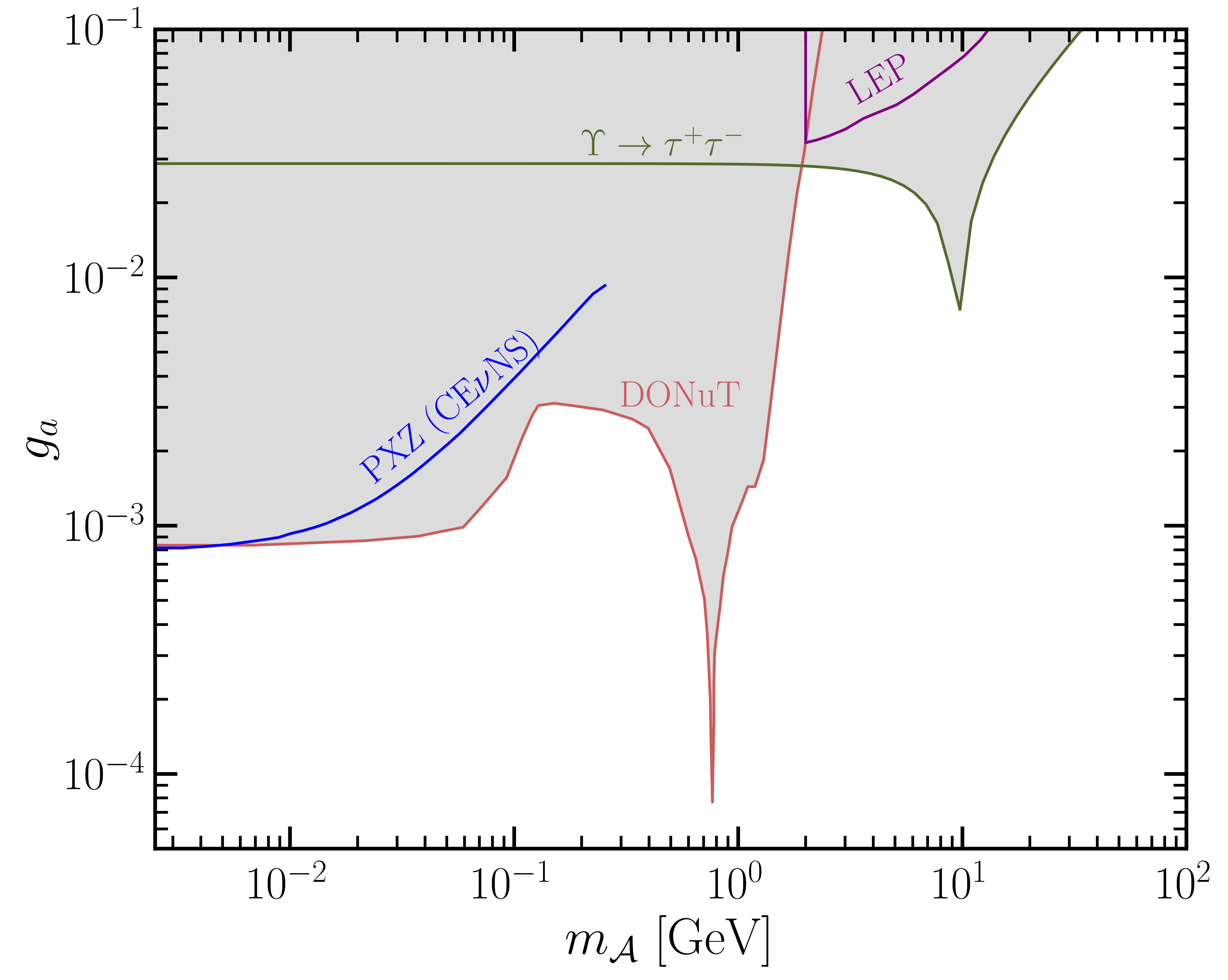}
        \label{fig:bounds_13_vis}
    }
    \caption{Bounds on (a) an invisibly decaying and (b) a visibly decaying $B-3L_{\tau}$ (model C) axial gauge boson.}
    \label{fig:bounds_13_combined}
\end{figure}

\medskip

\noindent {\textbf{Neutral meson to lepton-pair searches}}.
In the standard model, processes $\pi^0\to e^+e^-$ and $\eta\to \mu^+\mu^-$ are both loop and helicity suppressed. The calculation of these processes in the SM is dominated by virtual two-photon intermediate states. Both processes have been experimentally observed. For the first process, there was previously a $2-3~\sigma$ discrepancy between the theory (SM)  ~\cite{Dorokhov:2007bd, Masjuan:2015lca} and KTeV's experimental value~\cite{KTeV:2006pwx}, while for the second process, the theoretical and experimental values agree with each other. {The recent NA62 measurement~\cite{Rosa:2025irf, Husek:2024gnn}, however, found that the branching ratio measurements for $\pi^0\to e^+e^-$ process are in agreement with standard model calculations~\cite{Husek:2024gnn}. We therefore take the new contribution to be one sigma of the SM prediction.}
The branching fractions~\cite{ParticleDataGroup:2024cfk}, along with the $1\sigma$ errors, are,
\begin{equation}
    \begin{aligned}
        \text{BR}(\pi^0 \to e^+ e^-) &= (6.48 \pm 0.33)\times 10^{-8} \\
        \text{BR}(\eta \to \ \mu^+ \mu^-) &= (5.8 \pm 0.8) \times 10^{-6}
\end{aligned}
\end{equation}

In our model, there is a new nonzero tree-level contribution to $\pi^0\to e^+e^-$ decay mediated by the ${\cal A}$ field. To estimate its effect, we need to first write down the interaction of ${\cal A}$ with the mass eigenstate quarks and leptons. The isospin singlet part of the current does not couple to pion due to isospin conservation. The leading term that is an isospin vector and couples to pions is  given by: 
\begin{eqnarray}
{\cal L}_{{\cal A}}~=~g_a {\cal A}_\mu[(-\frac{4}{3\theta^2_{uR}}-\frac{2}{3\theta^2_{dR}})1/2(\bar{u}_R\gamma^\mu u_R-\bar{d}_R\gamma^\mu d_R)+\bar{e}\gamma^\mu\gamma_5 e]
\end{eqnarray}
where $\theta_{uR}\simeq \frac{h_uv_R}{f_uv_\eta}$ and 
$\theta_{dR}\simeq \frac{h_dv_R}{f_dv_\eta}$ and we choose $v_R=v_\eta$ as an example.
We would like to estimate the strength of this interaction allowed by the $\pi^0\to e^+e^-$ decay data. Following Refs.~\cite{Masjuan:2015lca} and ~\cite{Kahn:2016vjr}, we estimate the contribution of
${\cal A}$ in model A to $\pi^0\to e^+e^-$ and find it to be:
\begin{eqnarray}
    M^{\pi}_{\cal A}~=~-{2\sqrt{2}}f_\pi m_\pi m_e \epsilon\frac{g^2_a}{ M^2_{\cal A}}
\end{eqnarray}
where $\epsilon\simeq (-4/3 \theta^2_{uR}-2/3\theta^2_{dR}) {\simeq 10^{-3}}$ with $\theta_{uR}\simeq \frac{h_uv_R}{f_uv_\eta}$ and 
$\theta_{dR}\simeq \frac{h_dv_R}{f_dv_\eta}$. To calculate the new physics contribution, we included this term along with the Standard Model Matrix element given below,
\begin{eqnarray}
    M_{SM}~=~\frac{2\sqrt{2}m_e m_\pi \alpha^2}{\pi^2 f_\pi} A
\end{eqnarray}
where $A$ is a parameter defined in ~\cite{Dorokhov:2007bd},\cite{Masjuan:2015lca} and is estimated to be $A\simeq 10$. We then place exclusion bounds such that $\text{BR}(\pi^0\to e^+ e^-)_{\mathcal{A}} - \text{BR}(\pi^0\to e^+ e^-)_{\text{SM}} < 0.33\times 10^{-8}$. 

Similarly, $\eta\to \mu^+\mu^-$, which has also been experimentally observed, appears to be consistent with the standard model predictions with $1\sigma$ confidence.  To estimate the new contribution to this in our model, we first write down the relevant interaction for quarks, keeping the pure octet as the dominant contribution to $\eta\to \mu^+\mu^-$.
\begin{eqnarray}
    {\cal L}_\eta~=~g_a {\cal A}_\mu[(\bar{u}_R\gamma^\mu u_R+\bar{d}_R\gamma^\mu d_R-2\bar{s}_R\gamma^\mu s_R)(1/6-1/3\theta^2_{sR})]
\end{eqnarray}
where $\theta_{sR}\simeq (h_sv_R)/f_\eta v_\eta\simeq 10^{-1.5}$ for $v_R\sim v_\eta$. We have assumed the $SU(3)$ singlet part of $\eta$ meson to be small.
The contribution of ${\cal A}$ to $\eta\to \mu^+\mu^-$ can then be estimated as (following~\cite{Kahn:2016vjr}).
\begin{eqnarray}
  M^{\eta}_{\cal A}~=~\sqrt{\frac{8}{3}} m_\eta m_\mu \epsilon_\eta\frac{g^2_a}{ M^2_{\cal A}}{2\tilde{F}}  
\end{eqnarray}
where $\tilde{F}=113$ MeV~\cite{Kahn:2016vjr} is the effective $\eta$ decay to vacuum coupling and $\epsilon_\eta=1/2(1/3-2/3\theta^2_{sR})$. This is to be compared with the SM contribution~\cite{Kahn:2016vjr}
\begin{eqnarray}
    M_{SM}~=~{4\sqrt{2}m_\mu m_\eta \alpha^2} f_{\eta\to \gamma\gamma} A(m^2_\eta) 
\end{eqnarray}
with $f_{\eta\to \gamma\gamma}\simeq 0.274$ GeV$^{-1}$ and $A(m^2_\eta)\sim -1.5-i5.42$. This arises from the two $\gamma$ intermediate states as in the pion case. Similar to the $\pi^0 \to e^+ e^-$ scenario, we place exclusion bounds for $\text{BR}(\eta\to \mu^+ \mu^-)_{\mathcal{A}} - \text{BR}(\eta\to \mu^+ \mu^-)_{\text{SM}} < 0.8\times 10^{-6}$.

\noindent \textbf{Other constraints} We would like to emphasize further that constraints from other experiments, such as electron-flavored reactor neutrino E$\nu$ES ~\cite{Coloma:2022avw, TEXONO:2009knm}, measured muon $g-2$~\cite{Muong-2:2025xyk}, and SN 1987a~\cite{Chang:2016ntp}, are subdominant relative to the depicted bounds.  Therefore, we refrain from showing them.

\subsubsection*{Invisible-specific bounds}

\noindent \textbf{Missing energy searches.} The NA64e~\cite{NA64:2023wbi, NA64:2021xzo} and NA64$\mu$~\cite{NA64:2024nwj} experiments search for invisible mediators that couple to electrons and muons, respectively. They observe the deviation in the incoming lepton beam $l = e,\mu$, due to bremsstrahlung in the dark sector, such as $l^- +\mathcal{N} \to l^- + \mathcal{N}+\mathcal{A}$ for $m_{\cal A}< 1$ GeV. The lack of any deviation in the beam translates into constraints on the cross-section of lepton bremsstrahlung, as shown in the red and green lines in Fig.~\ref{fig:bounds_univ}. We would like to highlight that NA64e has the most stringent constraints in this model, particularly for $m_\mathcal{A} \lesssim 100~\text{MeV}$. We further note that the shape of the exclusion curves differs for NA64e and NA64$\mu$, due to the scale of electron and muon masses.

\medskip

\noindent \textbf{Collider experiments: Monophoton and Monojet searches}. Electron-positron colliders such as BaBar~\cite{BaBar:2008aby} and LEP~\cite{Hook:2010tw} constrain the mass and coupling of invisibly decaying $\mathcal{A}$ through monophoton searches from processes such as $e^+ +e^- \to \gamma + \mathcal{A} $, followed by $\mathcal{A} \to \rm{inv.}$. These constraints are shown in brown and purple contours in Fig.~\ref{fig:bounds_univ}, where the upper mass limit appears from the center of mass energy of the $e^+ e^-$ system, which is around $10$~GeV at BaBar. At LEP, we observe a resonance near the $Z$ boson, after which the limits become subdominant. The LHC also places a constraint on invisibly decaying axial gauge bosons through monojet searches~\cite{CMS:2021far, CMS:2023hwl, ATLAS:2021kxv} for $m_{\mathcal{A}} > 100~\rm{GeV}$. Following recent studies such as Ref.~\cite{Agashe:2024owh}, we extend them down to $100~\rm{MeV}$ masses.

\medskip 

\noindent \textbf{Other Constraints.} We also depict constraints from spin-dependent direct detection~\cite{SENSEI:2020dpa, Essig:2017kqs, XENON:2019gfn}, assuming that dark matter accounts for the relic abundance through thermal freeze-out interactions, mediated by the axial gauge boson $\mathcal{A}$. Additionally, we show constraints from the SQUID experiment~\cite{Graham:2013gfa} that looks for the dark Stodolsky effect~\cite{Rostagni:2023eic} of ambient dark matter assuming $m_{\zeta} = m_{\mathcal{A}}/3$.

{There are also constraints on the model from the upper limit on  $\Upsilon\to {\cal A}+\gamma$ decay width~\cite{BaBar:2010eww, ParticleDataGroup:2014cgo}. However, since a vector-like particle going to two vector particles is forbidden by Yang's theorem, the final state of ${\cal A}$ can only involve its longitudinal mode~\cite{Babu:2017olk}. The amplitude is then given by the operator $\bar{b}\gamma_\mu \gamma_5 bk_\mu/m_{\cal A}$ leading to the final branching ratio, from PDG values, being given by
\begin{equation}
    \frac{B(\Upsilon\to {\cal A}+\gamma)}{B(\Upsilon\to e^+e^-)} \sim 2\frac{g^2m^2_b}{e^2 m_{\mathcal{A}}^2} \leq \frac{4.5\times 10^{-6}}{0.0239}
\end{equation}
We, therefore, constrain the parameters based on the above inequality in Figs.~\ref{fig:bounds_univ} and~\ref{fig:bounds_13}.
}

\subsubsection*{Visible-specific bounds}

In the case where $\mathcal{A}$ does not promptly decay into dark matter, they can be observed at various experiments dominantly through their decays to $e^+ e^-$, $\mu^+ \mu^-$, and $\pi^+ \pi^-$. Additionally, their lifetime is determined by the decay widths to neutrinos, charged leptons, and hadrons (note that their decays to neutrinos are challenging to observe at many detectors). We summarize the various types of searches for visibly decaying $U(1)_a$ gauge bosons below.

\medskip

\noindent \textbf{Electron beam-dump experiments} Searches for new particles produced from the bremsstrahlung of electrons at Electron beam-dump experiments such as A1~\cite{Merkel:2014avp}, APEX~\cite{APEX:2011dww}, E141~\cite{Riordan:1987aw}, E137~\cite{Bjorken:1988as}, E774~\cite{Bross:1989mp}, KEK~\cite{Konaka:1986cb}, and Orsay~\cite{Davier:1989wz}, have searched for new particles that can be produced from electron bremsstrahlung, $e N\to e N \mathcal{A}$, which consequently decay to $e^+ e^-$ at the detector. These bounds severely constrain couplings between $10^{-9} \lesssim g_a \lesssim 10^{-3}$ for $m_{\mathcal{A}} \lesssim 100~\text{MeV}$ as shown in Fig.~\ref{fig:bounds_univ_vis}. 

\medskip

\noindent \textbf{Proton beam-dump experiments} Proton beam dump experiments set limit on new gauge bosons that can be produced from neutral meson decays, such as $\pi^0/\eta \to \gamma A'$ and proton bremsstrahlung $p N\to p N \mathcal{A}$. However, since a purely axial gauge boson doesn't mix with the SM photon, they cannot be produced from the decay of neutral pseudoscalar mesons. However, proton bremsstrahlung still serves as an efficient method to probe axial gauge bosons. Therefore constraints from NuCal's~\cite{Blumlein:1990ay, Blumlein:1991xh} searches for $\mathcal{A} \to e^+ e^-/\mu^+ \mu^-$ constrains our parameters space.

\medskip

\noindent\textbf{Collider experiments: Visible searches.} Electron-positron colliders such as BaBar~\cite{BaBar:2014zli} search for $e^+ e^- \to \mathcal{A} \gamma$. BaBar can search for them through their decays to $e^+ e^-$ final states, leading to stringent bounds to our model for $10~\text{MeV} \lesssim m_{\mathcal{A}} \lesssim 10~\text{GeV}$. We also include bounds from the LHCb~\cite{LHCb:2017trq} proton-proton collider searches for axial gauge bosons through various channels $u\bar u \to \mathcal{A}, d\bar d \to \mathcal{A}$, etc.,~\cite{Ilten:2018crw} set limits up to $m_{\mathcal{A}} \lesssim 100~\text{GeV}$.

\subsection{Axial $B-3L_\tau$ gauge model}

We depict the bounds on an invisible and a visible $B-3L_{\tau}$ axial gauge boson in Figs.~\ref{fig:bounds_13} and \ref{fig:bounds_13_vis}, respectively. Since the gauge boson couples only to third-generation fermions, many of the constraints involving electrons and muons do not apply here. These include E$\nu$ES, NA64e, NA64$\mu$, and $e^+e^-$ missing energy searches. Note that since the $\mathcal{A} -\gamma$ mixing is absent, these constraints do not apply even at the one-loop level. However, since a third of the solar neutrinos are tau-flavored, constraints from CE$\nu$NS searches at PandaX, XENONnT, and LZ would apply in our model. These bounds apply to both the visible and invisible $B-3L_{\tau}$ gauge bosons. Additionally, the monojet searches at the LHC apply to the invisible scenario as $\mathcal{A}$ couples to all quarks. 

Additionally, constraints from DONuT~\cite{DONuT:2007bsg}, which look for new sources of $\nu_\tau$s that can arise from the decay of new tau-philic mediators~\cite{Kling:2020iar}, apply to our model. These limits appear when the $\mathcal{A}$ is produced from proton bremsstrahlung $pN\to pN\mathcal{A}$, followed by their decay to tau-neutrinos. These are then detected by looking for $\nu_\tau$ charged-current interactions, leading to single-tau signatures at the detector. We note that there is a resonance at 750~MeV due to the enhanced production of $\mathcal{A}$ from $\rho-\omega$ resonance via proton bremsstrahlung. Since the event rate is proportional to $\text{BR}(\mathcal{A} \to \bar{\nu}_\tau \nu_\tau)$, the bounds are more stringent for the visible scenario where $\text{BR}(\mathcal{A} \to \bar{\nu}_\tau \nu_\tau)_{\text{vis}} \simeq 0.6$, whereas it is suppressed due to the presence of the strongly-coupled dark matter ($\text{BR}(\mathcal{A} \to \bar{\nu}_\tau \nu_\tau)_{\text{invis}} \simeq g_a^2/g_{\zeta}^2$). This leads to stronger constraints from DONuT in Fig.~\ref{fig:bounds_13_vis} than in Fig.~\ref{fig:bounds_13}.

Since $\mathcal{A}$ couples to $\tau$s and not $\mu$s, the constraints of the lepton universality of the $\Upsilon$ mesons constrain the parameters that result in $\text{BR}(\Upsilon \to \tau^+\tau^-)$ to be much different from those measured by BaBar~\cite{BaBar:2010esv}. The measured value is,
\begin{equation}
    R_{\mu\tau} =\frac{\text{BR}(\Upsilon \to \tau^+ \tau^-)}{\text{BR}(\Upsilon \to \mu^+ \mu^-)} = 1.005 \pm 0.013 \pm 0.022
\end{equation}
Since the additional diagram for decays to $\tau^+ \tau^-$ leads to~\cite{Babu:2017olk},

\begin{equation}
    R_{\mu\tau} = 1 - 2 \left(\frac{g_a}{e} \right)^2\frac{m_{\Upsilon}^2}{m_{\Upsilon}^2 - m_{\mathcal{A}}^2}
\end{equation}

We therefore exclude parameters that result in $R_{\mu\tau} - 1.005$ greater than the allowed values $1\sigma$. We also show the constraints from the constraints on the $Z$ boson decay width~\cite{Ma:1998dp} (such as $Z \to \tau^+ \tau^- \mathcal{A},~\nu_\tau \bar{\nu}_\tau \mathcal{A},~q \bar{q} \mathcal{A}$) for a visibly decaying $U(1)_{B-3L_\tau}$ axial vector in Fig.~\ref{fig:bounds_13_vis} which constrain coupling of $\mathcal{O}(10^{-2}-10^{-1})$ for masses above $\sim 2~\text{GeV}$.

\medskip

\medskip

\noindent \textbf{Non-Standard Neutrino Interactions.} Since ${\cal A}$ couples to all fermions, including neutrinos, the model will have non-standard neutrino interactions of the form $\frac{g^2_a}{q^2-M^2_{\cal A}} Q_\nu Q_f \bar{\nu}\gamma^\mu\nu \bar{f}\gamma_\mu \gamma_5 f$. The current limits on the strengths of these interactions have been discussed in Refs.~\cite{Coloma:2023ixt} and~\cite{Proceedings:2019qno}, and the most stringent limit on them is from the quark sector and is of the order $G_F$, provided the exchange boson has a heavy mass. Since in our model, the value $q^2$ exceeds the value $M^2_{\cal A}$, the analysis of~\cite{Abbaslu:2025fwt}, which provides the limit on the axial nonstandard neutrino interaction,
does not apply.

Similar to the axial $B-3L_{\tau}$ model, one can construct a $B-3L_{\mu}$ model as well, which contains similar bounds and allowed parameters as shown in Fig.~\ref{fig:bounds_13}. Additionally, constraints from COHERENT, NA64$\mu$, BaBar4$\mu$~\cite{BaBar:2016sci}, and CMS4$\mu$~\cite{CMS:2024jyb} would apply. We leave this model and its phenomenological implications for future studies.



\section{Comments} 
\label{sec:comments}
\begin{itemize}

\item Two of the models we have considered have family universal axial quantum numbers, but this is not required by anomaly cancellation. Different assignments of the $U(1)_a$ charge to different families can also lead to anomaly free models. For example,Model C has precisely such an assignment for leptons and is more suited for explaining the MiniBooNE low energy excess while remaining consistent with observational constraints.

\item Model A could be part of higher unification models: e.g. the higher gauge symmetry could be $SU(3)_{cL}\times SU(3)_{cR} \times SU(2)_{L} \times SU(2)_{R} \times U(1)_{aL}\times U(1)_{aR}$ which is a subgroup of the grand unification (GUT) group $SU(5)_L\times SU(5)_R$~\cite{Babu:2023dzz}. The mass scales of the model will, of course, be constrained to be different due to the requirement of gauge coupling unification. The singlet fermions $N_{L,R}$, however, will be outside the gauge group. The GUT theory makes the $U(1)_a$ quantum number family universal. The Higgs fields employed to implement symmetry breaking in a GUT model will be different (see, for example, \cite{Babu:2023dzz}). 

\end{itemize}

\section{Conclusion} 
\label{sec:conclusions}
In conclusion, we have constructed three UV complete models for a low mass axial gauge boson. They can explain small neutrino masses, include a dark matter candidate using both the gauge and Higgs portal and also provide solutions to the strong CP problem. A novel feature of one class of these models (model A) is that there is an upper limit on the magnitude of the $U(1)_a$ gauge coupling constant, arising from the fact that the SM Higgs doublet shares a $U(1)_a$ quantum number. 

We use Models B and C to delineate the allowed parameter space, which can be expressed straightforwardly in terms of the axial gauge coupling, $g_a$, and the axial-vector boson mass, $m_{\mathcal{A}}$. In contrast, for Model A, the parameter space cannot be uniquely characterized solely in terms of $g_a$ and $m_{\mathcal{A}}$, as additional parameters enter nontrivially into the phenomenology. All of these models can accommodate light neutrino masses. The parameter space allowed for models B and C is determined after including the relevant experimental constraints. The light mediators and dark matter can be probed at various ongoing high-intensity lower-energy beam (electron and proton) facilities. In particular, in a follow-up paper~\cite{DKM}, we show how the MiniBooNE low-energy excess can be explained in the model C based on the axial $B-3L_{\mu,\,\tau}$ symmetry, using either dark matter or neutrino scattering  off the nucleons.

\vspace{.2in}

\section*{Acknowledgement} We would like to thank Roni Harnik and Pedro Machado for insightful discussions. The work of BD and AK is supported by the U.S. DOE Grant DE-SC0010813.

\appendix

\section{Symmetry breaking mechanisms}

In this appendix, we sketch the Higgs potential for both models and look for their minima as is relevant to the strong CP problem for models A and B. We show that the phases of the Higgs field VEVs vanish, as would be required to solve the strong CP problem using parity.

\subsection{Model A}
\begin{eqnarray}
V(\chi_L, \chi_R, \eta_a)~=~-\sum_i \mu^2_i \phi^\dagger_i\phi_i~+\sum_{i,j}\lambda_{ij}\phi^\dagger_i\phi_i \phi^\dagger_j\phi_j~+m_0\eta^2_2\eta_1+m_1\eta^3_2\eta^*_3+ m_2\eta_1\eta^*_2\eta_3+h.c.
\end{eqnarray}
Here the fields $\phi_i$ stand for $\chi_{L,R}, \eta_{1,2,3}$ and $m_{0,1,2}$ are real parameters, due to parity symmetry under which $\eta_i\to \eta^*_i$. We will assume that the mass terms for $\chi_{L,R}$ break the left right symmetry softly. 
To find VEV phases at the minima for this potential, we parameterize 
$<\chi^0_{L,R}>=v_{L,R}/\sqrt{2}$ where the phases in these fields have been removed by a $SU(2)_L$ and $SU(2)_R$ transformation. We now parameterize the VEVs of the $\eta_i$ fields as $<\eta_1>=v_{\eta,1}e^{i\alpha_1} ;
<\eta_2>=v_{\eta,2}e^{-i\alpha_2}; <\eta_3>=v_{\eta,3}e^{-3i\alpha_3}
$
The minima conditions on the potential when we vary the phases implies that
\begin{eqnarray}
    \alpha_1+2\alpha_2=0; ~~ 3\alpha_2-\alpha_3=0;~ \alpha_1-\alpha_2+\alpha_3=0
\end{eqnarray}
Solving these equations, we find that the VEVs can be written as

\begin{eqnarray}
\langle\chi^0_{L,R}\rangle=v_{L,R}/\sqrt{2}; \quad \langle\eta_1\rangle=v_{\eta,1}e^{i\alpha_1} ; \quad 
\langle\eta_2\rangle=v_{\eta,2}e^{-i\alpha_1/2};  \quad \langle\eta_3\rangle=v_{\eta,3}e^{-3i\alpha_1/2}
\end{eqnarray}
It is clear from the above equation that the phases from $\eta$ vevs can be rotated away by a gauge $U(1)_a$ transformation so that all the vevs are real. Thus, all CP violation in the model resides in the Yukawa couplings.

\subsection{Model B}

\begin{eqnarray}
V(\chi_L, \chi_R, \eta_a)~=~-\sum_i \mu^2_i \phi^\dagger_i\phi_i~+\sum_{i,j}\lambda_{ij}\phi^\dagger_i\phi_i \phi^\dagger_j\phi_j~+m_1\eta^3_1\eta^*_2+h.c.
\end{eqnarray}
Here the fields $\phi_i$ stand for $\chi_{L,R}, \eta_{1,2}$ and $m_1$ is a real parameter due to parity symmetry.
We follow the same procedure as in model A to find the phases in the VEVs.  
\begin{eqnarray}
\langle \chi^0_{L,R} \rangle=v_{L,R}/\sqrt{2};\quad \langle \eta_1 \rangle=v_{\eta,1}e^{i\alpha_1};\quad
\langle\eta_2\rangle=v_{\eta,2}e^{-i\alpha_1/3}.
\end{eqnarray}
As in model A, these vacuum phases can be rotated away by a $U(1)_a$ transformation, preserving the vacuum state to be CP conserving, with all CP violation residing in the Yukawa couplings, thus solving the strong CP problem.

\subsection{Model $A'$} The scalar potential in this case is given by
\begin{eqnarray}
V(\chi_L, \chi_R, \eta_a)~=~-\sum_i \mu^2_i \phi^\dagger_i\phi_i~+\sum_{i,j}\lambda_{ij}\phi^\dagger_i\phi_i \phi^\dagger_j\phi_j~+ \beta_{\Delta}(\chi_L\chi_L\Delta_L+L\leftrightarrow R +h.c.
\end{eqnarray}
Here the fields $\phi_i$ stand for $\chi_{L,R}, \eta,\Delta_{L,R}$ with mass parameters breaking L-R symmetry. We can choose $\mu^2_{\Delta_{L,R}}$ positive and large so that the vevs of $\Delta_{L,R}$ respectively are given by $ \kappa_{L,R}= \frac{\beta_{\Delta} v^2_{L,R}}{\mu^2_{\Delta_{L,R}}}$. Choosing $\mu^2_{\Delta_{L,R}}$ appropriately, we can get small value for $\kappa_L$ which gives small masses to the left-handed neutrinos and a convenient mass for $\nu_R$'s via type II seesaw.

\bibliography{ref}

\end{document}